\documentstyle [12pt,amsfonts] {article}
\input epsf

\newcommand{\beq}{\begin{equation}}
\newcommand{\eeq}{\end{equation}}
\newcommand{\beqs}{\begin{eqnarray}}
\newcommand{\eeqs}{\end{eqnarray}}

\catcode`@=11
\@addtoreset{equation}{section}
\@addtoreset{equation}{subsection}
\def\theequation{\ifnum\value{section}=0 \arabic{equation}\ignorespaces
\else \ifnum\value{section}=-1 A.\arabic{equation}\ignorespaces
\else \ifnum\value{subsection}=0 \thesection.\arabic{equation}\ignorespaces
\else \thesection.\arabic{subsection}.\arabic{equation}\ignorespaces
                           \fi
                      \fi
                 \fi}
\catcode`@=12

\begin{document}

\def\thefootnote{\fnsymbol{footnote}}

\baselineskip 5.0mm

\vspace{4mm}

\begin{center}

{\Large \bf Tutte Polynomials and Related Asymptotic Limiting Functions for 
Recursive Families of Graphs}

\vspace{8mm}

\setcounter{footnote}{0} Shu-Chiuan Chang {\footnote{email:
shu-chiuan.chang@sunysb.edu}} and 
\setcounter{footnote}{6} Robert Shrock {\footnote{email: 
robert.shrock@sunysb.edu; paper submitted in connection with
a talk by R.S. at the Workshop on Tutte Polynomials, Centre de Recerca
Matem\`{a}tica (CRM), Universitat Aut\`{o}noma de Barcelona, Sept. 2001}}

\vspace{6mm}

C. N. Yang Institute for Theoretical Physics  \\
State University of New York       \\
Stony Brook, N. Y. 11794-3840  \\

\vspace{10mm}

{\bf Abstract}

\end{center}

We prove several theorems concerning Tutte polynomials $T(G,x,y)$ for recursive
families of graphs.  In addition to its interest in mathematics, the Tutte
polynomial is equivalent to an important function in statistical physics, the
Potts model partition function of the $q$-state Potts model, $Z(G,q,v)$, where
$v$ is a temperature-dependent variable. We determine the structure of the
Tutte polynomial for a cyclic clan graph $G[(K_r)_m,L=jn]$ comprised of a chain
of $m$ copies of the complete graph $K_r$ such that the linkage $L$ between
each successive pair of $K_r$'s is a join $jn$, and $r$ and $m$ are arbitrary.
The explicit calculation of the case $r=3$ (for arbitrary $m$) is presented.
The continuous accumulation set of the zeros of $Z$ in the limit $m \to \infty$
is considered.  Further, we present calculations of two special cases of Tutte
polynomials, namely, flow and reliability polynomials, for cyclic clan graphs
and discuss the respective continuous accumulation sets of their zeros in the
limit $m \to \infty$.  Special valuations of Tutte polynomials give
enumerations of spanning trees and acyclic orientations. Two theorems are
presented that determine the number of spanning trees on $G[(K_r)_m,jn]$ and
$G[(K_r)_m,id]$, where $L=id$ means that the identity linkage.  We report
calculations of the number of acyclic orientations for strips of the square
lattice and use these to obtain an improved lower bound on the exponential
growth rate of the number of these acyclic orientations.

\vspace{16mm}

\pagestyle{empty}
\newpage

\pagestyle{plain}
\pagenumbering{arabic}
\renewcommand{\thefootnote}{\arabic{footnote}}
\setcounter{footnote}{0}

\section{Introduction}

\subsection{Tutte Polynomial} 

The Tutte polynomial \cite{tutte1}-\cite{tutte5} (sometimes called the
dichromatic or Tutte/Whitney polynomial) contains much information about a
graph and includes a number of important functions as special cases.  Some
additional early papers and reviews are \cite{crapo}-\cite{boll}.  In this
paper we shall present a number of results on Tutte polynomials for recursive
families of graphs and on related asymptotic limiting functions in the limit
where the number of vertices on these graphs goes to infinity.  In this
introductory section we give some relevant definitions, notation, and 
previous relevant theorems. 

\medskip

{\bf Def.} \quad Let $G=(V,E)$ be a graph with vertex and
edge sets $V$ and $E$ and let the cardinality of these sets be denoted as $|V|$
and $|E|$, respectively.  One way of defining the Tutte polynomial of a graph 
$G$ is 
\beq
T(G,x,y)=\sum_{G^\prime \subseteq G} (x-1)^{k(G^\prime)-k(G)}
(y-1)^{c(G^\prime)}
\label{t}
\eeq
where $G^\prime=(V,E^\prime)$ is a spanning subgraph of $G$, i.e., $E^\prime
\subseteq E$, $k(G^\prime)$ denotes the number of connected components of
$G^\prime$, and $c(G^\prime)$ denotes the number of independent circuits
(i.e. the co-rank) in $G^\prime$, satisfying $c(G^\prime) =
|E^\prime|+k(G^\prime)-|V|$.  The first term in the summand of (\ref{t})
can equivalently be written $(x-1)^{r(G)-r(G^\prime)}$ where the rank $r(G)$ of
a graph $G$ is defined as $r(G)=|V|-k(G)$. We shall be interested in connected
graphs here, so that $k(G)=1$.

An equivalent way to define the Tutte polynomial is as follows: if $H$ is
obtained from a tree with $|E|$ edges by adding $\ell$ loops, then 
\beq
T(H,x,y)=x^{|E|}y^\ell
\label{tgxy}
\eeq
and if $e \in E$ is not a loop or a bridge (= isthmus or co-loop), then
$T(G,x,y)$ satisfies the deletion-contraction property
\beq
T(G,x,y) = T(G-e,x,y)+T(G/e,x,y)
\label{trel}
\eeq
where $G-e \equiv G \backslash e $ denotes the graph with the edge $e$ deleted
and $G/e$ denotes the graph contracted on this edge, i.e. the graph with the
edge $e$ deleted and the two vertices that it joined identified.  It is
straightforward to establish that the definitions (\ref{tgxy})-(\ref{trel}) and
(\ref{t}) are equivalent. The definition of the Tutte polynomial can be
generalized from graphs to matroids \cite{crapo} but we shall not need this
generalization here.

An elementary consequence of the definition (\ref{t}), is that for a
planar graph $G$, the Tutte polynomial satisfies the duality relation
\beq
T(G,x,y) = T(G^*,y,x)
\label{tuttedual}
\eeq
where $G^*$ is the (planar) dual to $G$.

\subsection{Recursive Graphs}

{\bf Def.} \quad A recursive graph $G_m$ is a graph that is constructed by
connecting $m$ copies of a subgraph $H$ together sequentially in some
prescribed manner.  

\medskip

One important class of recursive graphs is provided by strips of regular
lattices, such as the square lattice.  Let us envision such a strip as
extending in the longitudinal direction, along the $x$ axis, and having length
$L_x=m$ vertices, with width $L_y$ along a transverse ($y$) axis (where no
confusion should result between these axes and the $x,y$ variables of the Tutte
polynomial).  Let $P_{L_y}$ denote the path graph containing $L_y$ vertices,
denoted $v_1, v_2, ..., v_{L_y}$.  Formally, one can define the strip of the
square lattice by starting with the subgraph $H=P_{L_y}$ and specifying the
linkage $L$ that connects two successive copies of $H$ as the identity linkage,
such that vertex $v_i$ of the first $P_{L_y}$ is connected by an edge to vertex
$v_i^\prime$ of the next $P_{L_y}$ for $i=1,...,L_y$.  This linkage will be
denoted $L=id.$ Finally, one specifies how the longitudinal ends are treated.
Possibilities include free ends, a cyclic strip, such that one identifies the
subgraph $H_{m+1}$ with $H_1$, and a M\"obius strip, such that one identifies
$H_{m+1}$ with $T(H_1)$, where $T(H)$ means a twist, i.e., a reversal of
transverse orientation.  We shall denote the cyclic and M\"obius square-lattice
strips as $sq(L_y,L_x,FBC_y,PBC_x)$ and $sq(L_y, L_x,FBC_y,TPBC_x)$, where, 
in physics nomenclature, $FBC$ and $PBC$ refer to
free and periodic boundary conditions.  One may also consider $H=C_{L_y}$,
where $C_n$ denotes the circuit graph with $n$ vertices.  This is equivalent to
periodic transverse boundary conditions, $PBC_y$, in physics terminology.  Then
the square-lattice strip graphs $sq(L_y, L_x,PBC_y,FBC_x)$, $sq(L_y,
L_x,PBC_y,PBC_x)$, and $sq(L_y,L_x,PBC_y,TPBC_x)$ can be embedded
in cylindrical, torus, and Klein bottle surfaces, respectively.  For fixed
$L_y$, we shall refer to the set of strip graphs $sq(L_y,L_x,FBC_y,PBC_y)$ 
for variable $L_x$ as a family, and so forth for other
recursive families of graphs.  An early study of chromatic and Tutte
polynomials for recursive graphs is \cite{bds}. Some calculations of Tutte
polynomials for lattice strip graphs are in \cite{bcc}-\cite{ts}.

Another important class of recursive graphs is comprised of clan graphs
\cite{readcarib81}.  First, recall two auxiliary definitions: A complete graph
$K_r$ is a graph containing $r$ vertices with the property that each vertex is
connected by edges to every other vertex.  (Clearly, $|E|={r \choose 2}$ for
$K_r$.) The join of two graphs $H_1$ and $H_2$, denoted $H_1+H_2$, is the graph
formed by connecting each vertex of $H_1$ to all of the vertices of $H_2$ with
edges.  Then

\medskip

{\bf Def.} \quad   A clan graph is a
recursive graph composed of a set of $m$ complete graphs $K_{r_1}$,
$K_{r_2}$,..., $K_{r_m}$ such that the linkage between two adjacent pairs, say,
$K_{r_i}$ and $K_{r_{i+1}}$, is a join.  

\medskip

{\bf Def.} \quad A homogeneous clan graph is a clan graph with the property
that all of the $K_{r_i}$ are the same, say $K_r$.  

\medskip

{\bf Def.} \quad A cyclic clan graph of length $m$ is a clan graph with the
$K_{r_i}$'s arranged around a circle, i.e., with the identifications
$K_{r_{i+m}} = K_{r_i}$ (i.e., with periodic longitudinal boundary conditions,
in physics terminology).

\medskip 

{\bf Def.} \quad A homogeneous cyclic clan graph is thus a cyclic clan graph
for which $r_i = r \ \forall \ i$; we shall denote this as $G[(K_r)_m,jn]$,
where $jn$ indicates the join linkage between each successive pair of $K_r$'s.
We shall omit the qualifier ``homogeneous'' where it is obvious from the
context. 

Previously, we calculated the Tutte polynomial for the family of cyclic clan
graphs of length $m$ composed of $K_2$'s, i.e., $G[(K_2)_m,jn]$ \cite{ka}.
This family is equivalent to the family of cyclic strips of the square lattice
of width $L_y=2$ and length $L_x=m$ with next-nearest-neighbor spin-spin
interactions.  The asymptotic accumulation set of the zeros of the
corresponding Potts model partition function was determined in the limit $m \to
\infty$. 

We remark on some basic properties of the graph $G[(K_r)_m,jn]$. This has
\beq
|V|=mr \ , \quad |E|=\frac{mr(3r-1)}{2} 
\label{ve}
\eeq
and is a $\Delta$-regular graph with uniform vertex degree $\Delta=3r-1$ (here
a $\Delta$-regular graph is one in which all vertices have the same degree
$\Delta$).  It is straightforward to show that if one cuts the cyclic clan
graph $G[(K_r)_m,jn]$ at one of linkages between successive $K_r$'s, say
$K_{r_i}$ and $K_{r_{i+1}}$ and then reconnects the vertices of these two
complete graphs with each other with a twist, one obtains precisely the same
graph.  Recall that a clique of a graph $G$ is defined as the maximal complete
subgraph of $G$.  For $m$ sufficiently large as to avoid degenerate special
cases, the clique of $G[(K_r)_m,jn]$ is $K_{2r}$.

\subsection{Equivalence Between Tutte Polynomial and Potts Model Partition 
Function}

Besides its interest for graph theory, the Tutte polynomial has a close
connection with statistical physics, since it is, up to a prefactor, equal to
the partition function for a certain model of phase transitions and cooperative
phenomena known as the $q$-state Potts model \cite{wurev,potts,kf}.  We review
this connection here since we shall use it below. 

\medskip

{\bf Def.} \quad  For a
statistical mechanical system in thermal equilibrium at temperature $T$, the
partition function (= sum over all spin configurations $\{\sigma\}$ ) for the
$q$-state Potts model on the graph $G=(V,E)$ is given by
\beq
Z(G,q,v) = \sum_{ \{ \sigma \} } e^{-\beta {\cal H}}
\label{zfun}
\eeq
with the Hamiltonian describing the spin-spin interaction 
\beq
{\cal H} = -J \sum_{\langle i j \rangle} \delta_{\sigma_i \sigma_j}
\label{ham}
\eeq
where $\sigma_i=1,...,q$ are the effective spin variables on each vertex $i \in
V$; $J$ is the spin-spin coupling; $\beta = (k_BT)^{-1}$; $k_B$ is the
Boltzmann constant; and $\langle i j \rangle$ denotes the edge in $E$ joining
the vertices $i$ and $j$ in $V$.  

We use the notation 
\beq
K = \beta J \ , \quad a = e^K \ , \quad v = a-1
\label{kdef}
\eeq
(where $v$ should not be confused with $V$) so that the physical ranges are (i)
$J > 0$ and hence $a \ge 1$, i.e., $v \ge 0$ corresponding to $\infty \ge T \ge
0$ for the Potts ferromagnet, and (ii) $J < 0$ and hence $0 \le a \le 1$, i.e.,
$-1 \le v \le 0$, corresponding to $0 \le T \le \infty$ for the Potts
antiferromagnet.  An important function in physics is the (reduced) free energy
per site $f$:

\medskip

{\bf Def.} \quad The (reduced) free energy of the $q$-state Potts model on the
graph $G=(V,E)$ is 
\beq
f(\{G\},q,v) = \lim_{|V| \to \infty} \ln [ Z(G,q,v)^{1/|V|}] 
\label{ef}
\eeq
where we use the symbol $\{G\}$ to denote the formal limit $\lim_{|V| \to
\infty}G$ for a given family of graphs.  As discussed in \cite{a}, for
physically relevant values of $q$ and $v$, $Z(G,q,v)$ is positive and hence it
is clear which of the $1/|V|$'th roots (of which there are $|V|$) one should
choose in evaluating (\ref{ef}).  However, in other regions of the space of
$(q,v)$ variables, $Z(G,q,v)$ can be negative or complex, so that there is no
canonical choice of which root to take in (\ref{ef}) and only the quantity 
$|\exp(f(\{G\},q,v))|$ can be obtained unambiguously.

To show the equivalence of the Potts model partition function and Tutte
polynomial, we shall recall the elementary theorem \cite{kf}

\medskip

{\bf Theorem} \ \cite{kf} \quad The Potts model partition function
on the graph $G=(V,E)$ can be expressed as 
\beq
Z(G,q,v) = \sum_{G^\prime \subseteq G} q^{k(G^\prime)}v^{|E^\prime|}
\label{cluster}
\eeq
where $G^\prime$ is a spanning subgraph of $G$. 

\medskip

{\bf Proof}  This theorem is proved by observing that 
\beq
Z(G,q,v) = \sum_{ \{ \sigma \} } \prod_{\langle i j \rangle} (1 + 
v\delta_{\sigma_i \sigma_j})
\label{prodrel}
\eeq
and carrying out the indicated product over edges and sum over spin
configurations. $\Box$

\medskip

The formula (\ref{prodrel}) allows one to generalize $q$ from ${\mathbb Z}_+$
to ${\mathbb R}$; more generally, it allows one to generalize $q$ and $v$ to
${\mathbb C}$. 

One then has the well-known and important result 

\medskip

{\bf Corollary } \quad For a graph $G=(V,E)$ the Potts model
partition function $Z(G,q,v)$ is related to the Tutte polynomial $T(G,q,v)$
according to 
\beq
Z(G,q,v)=(x-1)^{k(G)}(y-1)^{|V|}T(G,x,y) 
\label{zt}
\eeq
where 
\beq
x=1+\frac{q}{v}
\label{xdef}
\eeq
and
\beq
y=a=v+1 
\label{ydef}
\eeq
so that 
\beq
q=(x-1)(y-1) \ .
\label{qxy}
\eeq

\medskip

{\bf Proof}  \quad  This follows immediately from (\ref{cluster}) and the 
definition (\ref{t}). $\Box$

\subsection{Chromatic, Flow, and Reliability Polynomials} 

An important property of the Tutte polynomial is the fact that it is a
Tutte-Gr\"othendieck (TG) invariant \cite{bryl,ow,welsh}.  Thus, consider a
function $f$ (not to be confused with the reduced free energy $f$ in
(\ref{ef})) that maps a graph $G$ to the elements of some field $K$.  Let $f_b
= f(bridge)$ and $f_\ell = f(loop)$. It will suffice to consider a connected
graph $G$; for disconnected graphs, one requires that $f(G_1 \cup G_2) =
f(G_1)f(G_2)$.

\medskip

{\bf Def.} \quad The function $f: \ G \to K$ is a Tutte-Gr\"othendieck
invariant if (1) if $e \in E$ is a bridge, then $f(G)=f_b f(G/e)$, (2) if $e
\in E$ is a loop, then $f(G)=f_\ell f(G/e)$, (3) if $e \in E$ is neither a
bridge nor a loop, then 
\beq
f(G) = af(G-e)+bf(G/e) \quad a,b \ne 0 \ . 
\label{dcrel}
\eeq

Next, we recall the 

\medskip 

{\bf Theorem } \ \cite{bo} 
\beq
f(G) = a^{|E|-|V|+1}b^{|V|-1}T(G,\frac{f_b}{b},\frac{f_\ell}{a}) \ . 
\label{tgtheorem}
\eeq
For the proof, see \cite{bo}.  As a consequence of this, many important
polynomial functions of graphs that are expressible in terms of TG invariants
are particular cases of the Tutte polynomial.  To render our discussion
self-contained, we review the relevant definitions and relations here.  

The first special case is the chromatic polynomial $P(G,q)$, which counts the
number of ways of coloring the vertices of $G$ subject to the constraint that
no adjacent pairs of vertices have the same color \cite{birk}-\cite{rtrev}.
Then
\beq
P(G,q) = (-q)^{k(G)}(-1)^{|V|}T(G,1-q,0) \ .
\label{pt}
\eeq
This result follows from (\ref{tgtheorem}) by observing that $q^{-1}P$ is a TG
invariant with $q^{-1}P(bridge) = q-1$, $q^{-1}P(loop)=0$, satisfying
(\ref{dcrel}) with $a=1$, $b=-1$.  A natural way for a physicist to prove this
is to observe first that the chromatic polynomial is identical to the $v=-1$
special case of the Potts model partition function: 
\beq
P(G,q) = Z(G,q,v=-1)
\label{pzrel}
\eeq
since for this value (corresponding to the zero-temperature Potts
antiferromagnet) no two adjacent spins can have the same value.  One then uses
(\ref{zt}) to obtain (\ref{pt}).  

Chromatic polynomials have been of interest in mathematics for many years since
Ref. \cite{birk}, owing in particular to their connection with the proper
face-coloring of bridgeless planar graphs and early efforts to prove the
four-color theorem, that a proper face-coloring of a bridgeless planar graph
$G$ can be accomplished using four colors.  This theorem is equivalent to the
statement that if $G$ is a loopless planar graph, then $P(G,4)$ is a positive
integer, i.e., there is a proper vertex coloring of $G$ with $q=4$ colors.
Chromatic polynomials are of interest for statistical physics because they are
equal to the partition function of the $q$-state Potts antiferromagnet at
zero-temperature (i.e., $v=-1$), as indicated in (\ref{pzrel}) above.  The
degeneracy of spin states per site for this $q$-state Potts antiferromagnet on
the $|V| \to \infty$ limit of the graph $G$ (usually a regular lattice with
some specified boundary conditions) is defined as
\beq
W(\{G\},q) = \lim_{|V| \to \infty} P(G,q)^{1/|V|} \ . 
\label{w}
\eeq
As discussed in \cite{w}, for the recursive graphs of interest here, for
sufficiently large real $q$, $P(G,q)$ is positive, so that one has a natural
choice for which of the $1/|V|$ roots to pick in evaluating (\ref{w}).  In
terms of the singular locus ${\cal B}$ to be defined below, one defines the 
region $R_1$ to be the maximal region in the $q$ plane to which one can
analytically continue $W(\{G\},q)$ from the interval $q > q_c(\{G\})$, where
$q_c(\{G\})$ denotes the maximal point where ${\cal B}$ intersects the real
axis \cite{w}.  For real $q > q_c$, and hence for all of region $R_1$, there is
no ambiguity in choosing (\ref{w}).  However, for regions $R_i$ not
analytically connected with $R_1$, only the magnitude $|W(\{G\},q)|$ can be
determined unambiguously since there is no canonical choice of which of the
$1/|V|$ roots to pick in evaluating (\ref{w}). The ground state entropy is
defined as
\beq
S_0(\{G\},q)=k_B \ln W(\{G\},q)
\label{s0}
\eeq
where $k_B$ is the Boltzmann constant.  The importance of this
is that the Potts antiferromagnet exhibits nonzero ground state entropy for
sufficiently large $q$ on a given $\{G\}$, an exception to the third law of
thermodynamics \cite{al}.  A physical system that exhibits nonzero residual
entropy at low temperatures is ice \cite{lp}. Previous studies of chromatic
polynomials include \cite{birk}-\cite{rtrev}, \cite{bds},
\cite{lieb}-\cite{kb56}. 

We next discuss the flow polynomial and start by recalling its definition. A
recent review is \cite{zhang}. 

\medskip

{\bf Def.} \quad Consider a connected graph $G=(V,E)$ with an orientation of
each edge $e \in E$, and assign to each edge a nonzero number in the
additive group ${\mathbb Z}_q/\{ 0 \}$, i.e., $1,2,...,q-1$.
A (nowhere-zero) flow on $G$ is an assignment of this type satisfying the 
property that the ingoing and outgoing flows are equal mod $q$ at each
vertex.  The number of such flows is given by the flow polynomial 
$F(G,q)$.

Remarks: \ As is implicit in the notation, $F(G,q)$ does not depend on the
orientation chosen for the edges.  Moreover, one can generalize the notion of a
flow to involve the assignment of an element of a nonzero element of any
abelian group of order $q$, not just ${\mathbb Z}_q$, to the edges of $G$, but
we shall not need this generalization here.  For an arbitrary bridgeless graph
it has been proved by Seymour that there is always a (nowhere-zero) 6-flow
\cite{seymour}, and it has been conjectured by Tutte that there is always a
(nowhere-zero) 5-flow \cite{tutte1}.

The flow polynomial $F(G,q)$ has the following properties. It obviously
vanishes on a bridge,
\beq
F(bridge,q)=0
\label{flowbridge}
\eeq
and satisfies
\beq
F(loop,q)=q-1 
\label{flowloop}
\eeq
\beq
F(G,q)=(q-1)F(G-e,q) \quad {\rm if} \ e \in E \ {\rm is \ a \ loop} \ . 
\label{floweloop}
\eeq
Further, 
\beq
F(G,q) = F(G/e,q)-F(G-e,q) \quad {\rm if} \ e \in E \ {\rm is \ not \ a \ loop}
 \ . 
\label{flowrel}
\eeq
Hence, from (\ref{tgtheorem}) the flow polynomial for a (connected) graph $G$
is given in terms of the Tutte polynomial by
\beq
F(G,q) = (-1)^{|E|-|V|+1}T(G,0,1-q) \ . 
\label{ft}
\eeq
An elementary result is that if $G$ is planar, then $P(G,q)=qF(G^*,q)$, where 
$G^*$ again denotes the planar dual of $G$.

Third, for a connected graph $G=(V,E)$, consider a related graph $H$ obtained
from $G$ by going through the full edge set of $G$ and, for each edge, randomly
retaining it with probability $p$ (thus deleting it with probability $1-p$),
where $0 \le p \le 1$.  The (all-terminal) reliability polynomial $R(G,p)$
gives the resultant probability that any two vertices in $G$ are connected,
i.e., that for any two such vertices, there is a path between them consisting
of a sequence of connected edges of $G$ \cite{colbourn}.  The reliability
polynomial of the connected graph $G$ is given in terms of the Tutte polynomial
by \cite{colbourn,welsh}
\beq
R(G,p) = p^{|V|-1}(1-p)^{|E|-|V|+1}T(G,1,1/(1-p)) \ .
\label{reliability}
\eeq
A basic property is that for $0 \le p \le 1$, it follows that $0 \le R(G,p) \le
1$.  Note that since, obviously, $R(G,1)=1$ for the connected graph $G$, it
follows that the prefactor $(1-p)^{|E|-|V|+1}$ in (\ref{reliability}) is always
cancelled by an inverse factor $(1-p)^{-|E|+|V|-1}$ from $T(G,1,1/(1-p))$ so
that $R(G,p)$ has no $(1-p)$ factor.  Studies of roots of the reliability
polynomial include \cite{bc,wagner,sokalzero}.

Special cases of Tutte polynomials also yield, up to prefactors, other
important functions, such as the Jones polynomial $V_L(t)$ for alternating
knots and links, obtained by setting $x=t$ and $y=-1/t$ \cite{th} (see, e.g. 
\cite{wuknot,jz}). 

\subsection{Special Valuations of Tutte Polynomials} 

For certain special values of the arguments $x$ and $y$, the Tutte polynomial
$T(G,x,y)$ yields various quantities of basic graph-theoretic interest
\cite{tutte3}-\cite{boll}.  We recall that a tree is a connected graph with no
cycles; a forest is a graph containing one or more trees; and a spanning tree
is a spanning subgraph that is a tree.  We also recall that the graphs $G$ that
we consider are connected.  A valuation that will be of interest below is the
number of spanning trees of a graph $G$, denoted by $N_{ST}(G)$.  From the
definition (\ref{t}), it is immediately evident that this is given by 
\beq 
N_{ST}(G)=T(G,1,1) \ .
\label{t11}
\eeq

A second valuation of the Tutte polynomial that will be of interest here yields
the number of acyclic orientations.  Thus, consider a connected graph $G$ and
define an orientation of $G$ by assigning a direction to each edge $e \in
E$. Clearly, there are $2^{|E|}$ of these orientations.  

\medskip

{\bf Def.} \quad  For a connected graph $G$, an acyclic orientation
is defined as an orientation that does not contain any directed cycles. (Here a
directed cycle is a cycle in which, as one travels along the cycle, all of the
oriented edges have the same direction.)  The number of such acyclic
orientations is denoted $a(G)$. 

A basic result, due to Stanley, is \cite{stanley} 
\beq
a(G)=(-1)^{n(G)}P(G,q=-1) \ .
\label{ap}
\eeq
Equivalently, in terms of the Tutte polynomial,
\beq
a(G)= T(G,x=2,y=0) \ .
\label{at}
\eeq
Other valuations include the number of spanning forests, 
$N_{SF}(G)=T(G,2,1)$, the number of connected spanning subgraphs, 
$N_{CSSG}(G)=T(G,1,2)$, and the number of spanning subgraphs, 
$N_{SSG}(G)=T(G,2,2)=2^{|E|}$, where the latter relation holds since $T(G,2,2)$
is the number of spanning subgraphs of $G$, and in counting these, one has the
two-fold choice for each edge of whether to include it in $G^\prime$ or to 
exclude it from $G^\prime$.

\subsection{Form of Tutte Polynomials for Recursive Families of Graphs}

In \cite{a}, the following theorem was proved: 

\medskip

{\bf Theorem 1} \cite{a} \quad Let $G_m$ be a recursive graph.  Then the Tutte
polynomial and Potts model partition function have the general forms
\beq 
T(G_m,x,y) = \sum_{j=1}^{N_{T,G,\lambda}} c_{T,G,j}(\lambda_{T,G,j})^m
\label{tgsum}
\eeq
\beq
Z(G_m,q,v) = \sum_{j=1}^{N_{Z,G,\lambda}} c_{Z,G,j}(\lambda_{Z,G,j})^m
\label{zgsum}
\eeq
where
\beq
N_{T,G,\lambda} = N_{Z,G,\lambda} \ .
\label{nttot}
\eeq
The coefficients $c_{T,G,j}$ and terms $\lambda_{T,G,j}$ in
eq. (\ref{tgsum}) and the coefficients $c_{Z,G,j}$ and terms 
$\lambda_{Z,G,j}$ in eq. (\ref{zgsum}) do not depend on the length $m$ of the
recursive graph. The coefficients $c_{Z,G,j}$ only depend on $q$, and 
satisfy 
\beq
c_{T,G,j} = \frac{\bar c_{T,G,j}}{x-1}
\label{ctfactor}
\eeq
with
\beq
\bar c_{T,G,j} = c_{Z,G,j} \ .
\label{ctcz}
\eeq
See \cite{a} for further discussion. 

Remark: \ A related property, shown in \cite{bds}, is that the Tutte
polynomials $T(G_m,x,y)$ for recursive graphs of different lengths $m$ satisfy
a finite recursion relation.  The special case of (\ref{tgsum}) for the
chromatic polynomial was noted earlier in \cite{bkw}, namely that for a
recursive graph $G_m$, 
\beq
P(G_m,q) = \sum_{j=1}^{N_{P,G,\lambda}} c_{P,G,j}(\lambda_{P,G,j})^m \ .
\label{pgsum}
\eeq

We next define the sums of coefficients $c_{Z,G,j}$. 

\medskip

{\bf Def.} \quad Let $G_m$ be a recursive graph so that the structural theorems
above hold.  Then
\beqs
C_Z(G) & = & \sum_{j=1}^{N_{Z,G,\lambda}} c_{Z,G,j} \cr\cr
       & = & \sum_{j=1}^{N_{T,G,\lambda}} \bar c_{T,G,j} \cr\cr
       & = & \bar C_T(G)
\label{ctzsum}
\eeqs
and
\beq
C_P(G) = \sum_{j=1}^{N_{P,G,\lambda}} c_{P,G,j} \ .
\label{cpsum}
\eeq
It was shown earlier how these sums are related to colorings of the repeated
subgraph \cite{cf}.

Before proceeding with our new results, we recall the standard notation from
combinatorics for the falling factorial $q_{(r)}$ and rising factorial
$q^{(r)}$
\beq
q_{(r)} = \prod_{s=0}^{r-1} (q-s)
\label{fallingfac}
\eeq
\beq
q^{(r)} = \prod_{s=0}^{r-1} (q+s) \ .
\label{risingfac}
\eeq
These satisfy the relations 
\beq
q^{(r)}=(q+r-1)_{(r)}
\label{risefall}
\eeq
\beq
q_{(r)}= r! {q \choose r}
\label{ffrel}
\eeq
\beq
q^{(r)}=r!{q+r-1 \choose r} = r! \sum_{j=1}^r {q \choose j} {r-1 \choose j-1}
\label{friserel}
\eeq
\beq
\frac{q_{(r+t)}}{q_{(r)}} = (q-r)_{(t)} 
\label{ffratio}
\eeq
\beq
r_{(r)}=r_{(r-1)}=r! \ . 
\label{rrel}
\eeq

\subsection{Continuous Accumulation Sets of Zeros}

In the context of cyclic clan graphs, we shall sometimes use the symbol
$G[(K_r)_\infty,jn]$ for the formal limit of the family as $m \to \infty$.
Since $T(G,x,y)$ is equal, up to a prefactor, to $Z(G,q,v)$, we shall usually,
with no loss of generality, consider the zeros in the ${\mathbb C}^2$ space
spanned by the Pott model variables $(q,v)$ rather than the zeros in the
${\mathbb C}^2$ space spanned by the Tutte variables $(x,y)$.

\medskip

{\bf Def.} \quad  Consider a recursive graph $G_m$ and the limit $m \to
\infty$.  The continuous accumulation set of zeros of $Z(G_m,q,v)$ as $m \to
\infty$ is denoted ${\cal B}$.  The slices of this locus in the $q$ plane for
fixed $v$ and in the $v$ plane for fixed $q$ will be denoted ${\cal B}_q$ 
and ${\cal B}_v$, respectively.  

Remarks: In cases where the locus ${\cal B}$ is nontrivial, it forms by the
merging together of the zeros as $m \to \infty$.  This continous locus may be
empty if the zeros accumulate at one or more discrete points.  A simple example
of the latter is provided by the special case of the chromatic polynomial of
the tree graph, $Z(T_m,q,-1)=P(T_m,q)=q(q-1)^{m-1}$.  Evidently, this
polynomial has only the discrete zeros $q=0$ with multiplicity one and $q=1$
with multiplicity $m-1$, so that there is no continuous accumulation set; 
${\cal B}_q=\emptyset$.  A simple example of a nontrivial locus is provided by
the Potts model partition function of the circuit graph.  An elementary
calculation yields $Z(C_m,q,v)=(q+v)^m+(q-1)v^m$, or equivalently,
$T(C_m,x,y)=y+\sum_{j=1}^{m-1} x^j$.  The locus ${\cal B}$ is given by the
solution to the equation $|q+v|=|v|$.  For fixed $v$, the locus ${\cal B}_q$ is
the circle centered at $q=-v$ with radius $|v|$. For fixed real $q$, the locus
${\cal B}_v$ is the vertical line ${\rm Re}(v)=-q/2$.  For the corresponding 
Tutte polynomial, ${\cal B}$ is the solution of the equation $|x|=1$, so that 
${\cal B}_x$, in an obvious notation, is the unit circle in the $x$ plane and
there is only a discrete zero in the $y$-plane, so that 
${\cal B}_y=\emptyset$. 

\medskip

{\bf Def.} \quad Consider a recursive graph $G_m$, so that $Z(G_m,q,v)$ has the
form (\ref{zgsum}), and let $q$ and $v$ be given.  Among the terms
$\lambda_{Z,G,j}$ in (\ref{zgsum}) the one with the maximal magnitude
$|\lambda_{Z,G,j}|$ at this point $(q,v)$ will be denoted as ``dominant'',
labelled $\lambda_{Z,G,dom.}$. 	This corresponds to the dominant term
$\lambda_{T,G,j}$ in (\ref{tgsum}) at the corresponding point $(x,y)$.

Remark: Since the $\lambda_{Z,G,j}$ are distinct, for a generic $(q,v)$, only
one term $\lambda_{Z,G,j}$ will be dominant.  As $m \to \infty$, the reduced
free energy will thus be determined only by this dominant term:
\beq
f=\ln [(\lambda_{Z,G,dom.})^{1/t}]
\label{fdom}
\eeq
where
\beq
t = \lim_{m \to \infty} \frac{|V|}{m} \ .
\label{texp}
\eeq
As one moves to another point $(q^\prime,v^\prime)$, it may happen that there
is a change in the dominant $\lambda$, from $\lambda_{Z,G,dom.}$ to, say,
$\lambda_{Z,G,dom}^\prime$.  If, indeed, this happens, then there is a
resultant nonanalytic change in the free energy $f$ as it switches from being
determined by $\lambda_{Z,G,dom.}$ to being determined by
$\lambda_{Z,G,dom.}^\prime$.  Hence

\medskip

{\bf Theorem 2} \quad  Consider a recursive family of graphs $G_m$ and let $m
 \to \infty$.  Then ${\cal B}$ is determined as the solution to the equation 
\beq 
|\lambda_{Z,G,dom.}|=|\lambda_{Z,G,dom.}^\prime| \ .
\label{degeneq}
\eeq
A corollary is that although $f(\{G\},q,v)$ is nonanalytic across ${\cal B}$, 
the quantity $|\exp(f(\{G\},q,v))|$ is continuous across this locus.  

Thus, for the Potts model partition function on a recursive family of graphs
$G_m$, in the $m \to \infty$ limit, the resultant locus ${\cal B}$ forms a
hypersurface in the ${\mathbb C}^2$ space spanned by $(q,v)$, and the reduced
free energy is nonanalytic on this surface. One can determine the corresponding
slices ${\cal B}_q$ in the $q$ plane for fixed $v$ and ${\cal B}_v$ in the $v$
(or equivalently, $a$) plane for fixed $q$.  In our earlier calculations of
Tutte polynomials and Potts model partition functions for recursive families of
graphs this program was carried out \cite{a}-\cite{ts}.

Similarly, for the one-variable specializations of the Tutte polynomial, we
denote the corresponding continuous accumulation set of zeros in the respective
variables as ${\cal B}_q$ for the $m \to \infty$ limits of the chromatic and
flow polynomials $P(G_m,q)$ and $F(G_m,q)$, and ${\cal B}_p$ for the $m \to
\infty$ limit of the reliability polynomial $R(G_m,p)$.  For a given $\{G\}$,
as one crosses the locus ${\cal B}_q$, there is a nonanalytic change in the
form of the function $W(\{G\},q)$ associated with a switch in the dominant term
$\lambda_{P,G,dom.}$.  Similar remarks hold for the flow and reliability
polynomials.

As a consequence of the fact that the Tutte polynomial has real coefficients
and of the relations connecting the chromatic, flow, and reliability
polynomials with the Tutte polynomial, it follows immediately that the set of
zeros of each of these polynomials for any $m$ in the respective complex planes
is invariant under complex conjugation, and, in the limit $m \to \infty$, the
respective continuous accumulation sets ${\cal B}$ are also invariant under
complex conjugation.  

Since for $q \in {\mathbb Z}_+$ and for sufficiently large $v$, $Z(G_m,q,v)$
grows exponentially rapidly as $m \to \infty$, as can be seen from the
structural theorem (\ref{zgsum}), the limit in eq. (\ref{ef}) defines a finite
reduced free energy $f(\{G\},q,v)$ as $|V| \to \infty$.  Similarly, for
sufficiently large positive $q$, $P(G_m,q)$ grows exponentially as $m \to
\infty$, and the definition in eq. (\ref{w}) yields a finite $W(\{G\},q)$.

Similar remarks hold for the flow polynomial $F(G_m,q)$ and reliability
polynomial, and we are thus led to introduce the definitions 

\medskip

{\bf Def.}  \quad Let $G_m$ be a recursive family of graphs and consider the
limit $m \to \infty$.  For sufficiently large real $q$, $F(G_m,q)$ is real and
positive, and we define
\beq
\phi(\{G\},q) = \lim_{|V| \to \infty} F(G_m,q)^{1/|V|}
\label{phi}
\eeq
with the canonical choice of $1/|V|$th root so that $\phi(\{G\},q)$ is real and
positive.  The function $\phi(\{G\},q)$ measures the number of (nowhere-zero)
$q$-flows per vertex in this limit. 

\medskip

{\bf Def.}  \quad Let $G_m$ be a recursive family of graphs and consider the
limit $m \to \infty$.  For $0 \le p \le 1$, we define 
\beq
\rho(\{G\},p) = \lim_{|V| \to \infty} R(G_m,p)^{1/|V|} \ . 
\label{rho}
\eeq
The function $\rho(\{G\},p)$ is a measure of the connectivity, normalized to be
per vertex, in this limit.

\subsection{Chromatic Polynomial for Cyclic Clan Graphs}

Since this paper includes new results on the Tutte polynomial of cyclic clan
graphs, it is useful to recall that the special case of the chromatic 
polynomial for the clan graph $G[(K_r)_m,jn]$ was calculated by 
Read \cite{readcarib81}; in our present notation, it is 
\beq
P(G[(K_r)_m,jn],q) = \sum_{d=0}^{r} \mu_d (\lambda_{P,r,d})^m
\label{pclan}
\eeq
where
\beq
\lambda_{P,r,d}=(-1)^d r_{(d)} \frac{q_{(2r)}}{q_{(r+d)}} = 
(-1)^d r_{(d)} (q-r-d)_{(r-d)}
\label{lamchrom}
\eeq
and the coefficient $\mu_d$ is a polynomial in $q$ of degree $d$ given by 
\beq
\mu_0=1
\label{mud0}
\eeq
\beq
\mu_d = {q \choose d} - {q \choose d-1} = \frac{q_{(d-1)}(q-2d+1)}{d!} 
\quad {\rm for} \quad 1 \le d \le r \ .
\label{mud}
\eeq
Note that
\beq
\lambda_{P,r,0}= (q-r)_{(r)}
\label{lamr0}
\eeq
\beq
\lambda_{P,r,r}=(-1)^r r!
\label{lamrr}
\eeq
and
\beq
\lambda_{P,r,r-1}=(-1)^{r-1}r!(q-2r+1) \ .
\label{lamrrminus1}
\eeq
The first few coefficients $\mu_d$ beyond $\mu_0$ are $\mu_1=q-1$,
$\mu_2=(1/2)q(q-3)$, $\mu_3=(1/3!)q(q-1)(q-5)$, and
$\mu_4=(1/4!)q(q-1)(q-2)(q-7)$.  For our discussion below it will be convenient
to extract the factor of $q$ that is present in $\mu_d$ for $d \ge 2$ and
define
\beq
\bar\mu_d = q^{-1} \mu_d \quad {\rm for} \ \ d \ge 2 \ .
\label{mudbar}
\eeq

The sum of the coefficients is 
\beq
C_P(G[(K_r)_m,jn]) = {q \choose r} \ .
\label{csumkr}
\eeq
Parenthetically, we note that Read actually proved a more general result,
namely the chromatic polynomial for an inhomogeneous cyclic clan graph where
the $r_1$, $r_2$,..., $r_m$ are, in general, different \cite{readcarib81}.
This chromatic polynomial does not have the form (\ref{pgsum}) since the
general inhomogeneous cyclic clan graph is not a recursive graph.  Biggs and
coworkers have also obtained a number of results on chromatic polynomials of 
recursive families of graphs comprised of $m$ copies of $K_r$ for arbitrary
linkage $L$, $G[(K_r)_m,L]$ \cite{lse9908,cprsg,amcp1}.  These are clan 
graphs if and only if the linkage is a join, $L=jn$. 

\subsection{Organization}

Having completed the discussion of relevant definitions and background, we now
proceed to present our new results. In next section, we shall determine the
structure of the Tutte polynomial for the cyclic clan graph $G[(K_r)_m,jn]$.
These general results will be illustrated by our previous calculation of the
Tutte polynomial for the case $r=2$, and here we shall present a calculation
for the case $r=3$ (for arbitrary $m$).  For the consideration of the limit $m
\to \infty$, it is convenient to work with the equivalent Potts model partition
function $Z(G[(K_3)_m,jn]$, and we shall give some results for the ${\cal B}$
for this function. Next, we present calculations of two special cases of Tutte
polynomials, namely, flow and reliability polynomials, for cyclic clan graphs
and discuss the respective continuous accumulation sets of their zeros in the
limit $m \to \infty$.  We then prove two theorems that determine the number of
spanning trees on $G[(K_r)_m,jn]$ and $G[(K_r)_m,id]$. Finally, we report
calculations of the number of acyclic orientations for strips of the square
lattice and use these to obtain an improved lower bound on the exponential 
growth rate of the number of these acyclic orientations.

\section{Structural Relations for Tutte Polynomials of $G[(K_r)_m,jn]$}

In \cite{cf}, relations were proved connecting the coefficients $\bar
c_{T,G,j}=c_{Z,G,j}$ in the Tutte polynomial and Potts model partition function
for a recursive graph $G$ with the corresponding coefficients in the chromatic
polynomial $P(G,q)$.  For sufficiently large integral $q$ these coefficients
are determined by the multiplicities of the the terms $\lambda_{T,G,j}$ or
equivalently $\lambda_{Z,G,j}$ as eigenvalues of an appropriately defined
coloring matrix (for the chromatic polynomial case, see \cite{matmeth}). From
this correspondence, it follows that the types of coefficients that enter in
the chromatic polynomial are the same as those that enter in the Tutte
polynomial.  Then 

\medskip

{\bf Lemma 1} \quad 
\beq
T(G[(K_r)_m,jn],x,y) = \frac{1}{x-1}\sum_{d=0}^r \mu_d \sum_{j=1}^{n_T(r,d)} 
(\lambda_{T,r,d,j})^m \ .
\label{tgsumclan}
\eeq
Equivalently, 
\beq
Z(G[(K_r)_m,jn],q,v) = \sum_{d=0}^r \mu_d \sum_{j=1}^{n_T(r,d)}
(\lambda_{Z,r,d,j})^m \ .
\label{zgsumclan}
\eeq

\medskip

{\bf Proof} \quad  This follows from eq. (\ref{ctcz}) together with 
(\ref{pclan}). $\Box$. 

Using (\ref{qxy}), (\ref{mud}), and (\ref{mudbar}), one can re-express 
(\ref{tgsumclan}) as 
\beq
T(G[(K_r)_m,jn],x,y) = \frac{1}{x-1}\sum_{d=0,1} \mu_d 
\sum_{j=1}^{n_T(r,d)}(\lambda_{T,r,d,j})^m
+ (y-1)\sum_{d=2}^r \bar\mu_d \sum_{j=1}^{n_T(r,d)}(\lambda_{T,r,d,j})^m
\label{tgsumclanexplicit}
\eeq
where the second sum is understood to be absent for $r < 2$. 

We next observe that 

\medskip

{\bf Lemma 2} 
\beq
C_Z(G[(K_r)_m,jn]) = \bar C_T(G[(K_r)_m,jn])) = \frac{q^{(r)}}{r!} \ .
\label{ctzsumclan}
\eeq

\medskip

{\bf Proof} \ Because the linkage between each repeated $K_r$ is a join, one
can freely permute the labels on the vertices of each $K_r$.  From the sum over
states in $Z(G,q,v)$ together with the relation (\ref{zt}), it follows that
the total multiplicity of colorings is enumerated by assigning colors freely to
each vertex of a given $K_r$, taking into account the above equivalences. But
this number is precisely $q^{(r)}/r!$. For example, for the case $r=3$ and
$q=3$, one has the colorings (111), (222), (333), (112), (113), (221), (223),
(331), (332), and (123).  We note the equivalences $(112) = (121) = (211)$ and
so for for colorings with two repeated colors, and $(123)=h(123)$ where $h$ is
an element of the permutation group $S_3$.  Hence, the total number of
inequivalent colorings is $10=3^{(3)}/3!$.  $\Box$

We now present our structure theorem: 

\medskip

{\bf Theorem 3} \quad  For $G[(K_r)_m,jn]$ the numbers $n_T(r,d)=n_Z(r,d)$, $0
\le d \le r$, are determined by 
\beq
n_T(r,r)=1
\label{ntrr}
\eeq
\beq
n_T(r,0)=n_T(r,1)=2^{r-1}
\label{ntr0}
\eeq
together with the recursion relation
\beq
n_T(r+1,d) = n_T(r,d)+n_T(r,d-1) \quad {\rm for} \ \ 2 \le d \le r \ .
\label{ntrecursion}
\eeq
For $d > r$, $n_T(r,d)=0$.

\medskip

{\bf Proof}  \quad From (\ref{ctzsumclan}) we have 
\beq
\bar C_T(G[(K_r)_m,jn])) = \sum_{d=0}^r n_T(r,d)\mu_d = \frac{q^{(r)}}{r!}
\ .
\label{ctsumclan}
\eeq
We differentiate this equation $r$ times to get $r+1$ linear equations for 
the $r+1$ unknowns $n_T(r,d)$, $d=0,1,...,r$.  Solving this equation yields the
results in the theorem. $\Box$ 

\medskip

{\bf Corollary 1} \quad For $G[(K_r)_m,jn]$, 
\beqs
n_T(r,d) & = & 2^{r-1} - \sum_{j=1}^{d-1} {r-1 \choose j-1} \cr\cr
         & = & \sum_{j=1}^{r-d+1} {r-1 \choose j-1}
\label{ntrdgen}
\eeqs

\medskip

{\bf Proof} \quad  These results follow in a straightforward manner by explicit
solution of the set of $r+1$ linear equations used in the proof of Theorem 3. 
$\Box$

For example, one has the special cases 
\beq
n_T(r,r-1)=r
\label{trrminus1}
\eeq
\beq
n_T(r,2)=2^{r-1}-1 \quad {\rm for} \ \ r \ge 2
\label{ntr2}
\eeq
\beq
n_T(r,3)=2^{r-1}-r \quad {\rm for} \ \ r \ge 3
\label{ntr3}
\eeq
and so forth. 

In Table 1 we show the numbers $n_T(r,d)$ and $N_{T,r,\lambda}$ 
for $1 \le r \le 8$. 

Defining $N_{T,r,\lambda}$ as 
\beq
N_{T,r,\lambda} = \sum_{d=0}^r n_T(r,d)
\label{nttotclandef}
\eeq
we find 

\medskip

{\bf Theorem 4} 
\beq
N_{T,r,\lambda} = (r+3)2^{r-2} \ .
\label{nttotclan}
\eeq

\medskip

{\bf Proof} \quad From (\ref{ntr0}) and the recursion relation
(\ref{ntrecursion}) for the set of coefficients $n_T(r,d)$ with a given $r$
(the $r$'th row of Table 1), it follows that
\beq
\sum_{d=1}^r n_T(r,d) = 2N_{T,r-1,\lambda}-2^{r-2} \ .
\label{rowsumright}
\eeq
Adding the final term $n_T(r,0)$ and using (\ref{ntr0}), we have 
\beq
N_{T,r,\lambda}=2N_{T,r-1,\lambda}+2^{r-2} \ .
\label{nttotrecursion}
\eeq
Combining this recursion relation for $N_{T,r,\lambda}$ with the initial value
$N_{T,1,\lambda}=2$ yields the theorem (\ref{nttotclan}). $\Box$ 

Remark: From our explicit calculation of $T(G[(K_3)_m,jn],x,y)$, we have found
that, in contrast to the situation for cyclic strips \cite{cf} and related
self-dual planar strip graphs \cite{dg,sdg}, for a given $r$, some
$\lambda_{T,r,d}$ with different $d$ may be equal to each other.  This does not
happen for $r=1$ or $r=2$; for $r=3$, we find (see eq. (\ref{lamtut301}) below)
that $\lambda_{T,3,0,1}=\lambda_{T,3,2,3}$.  Hence the total number of distinct
$\lambda_{T,r,d,j}$'s for $r=3$ is 11.

\begin{table}
\caption{\footnotesize{Table of numbers $n_T(r,d)$ and their sums,
$N_{T,r,\lambda}$ for $G[(K_r)_m,jn]$. Blank entries are zero.}}
\begin{center}
\begin{tabular}{|c|c|c|c|c|c|c|c|c|c|c|}
\hline\hline
$r \ \downarrow$ \ \ $d \ \rightarrow$
   & 0 & 1   & 2   & 3   & 4   & 5  & 6  & 7 & 8 & $N_{T,r,\lambda}$
\\ \hline\hline
1  & 1   & 1   &     &     &     &    &    &   &   &     2   \\ \hline
2  & 2   & 2   & 1   &     &     &    &    &   &   &     5   \\ \hline
3  & 4   & 4   & 3   & 1   &     &    &    &   &   &    12   \\ \hline
4  & 8   & 8   & 7   & 4   & 1   &    &    &   &   &    28   \\ \hline
5  & 16  & 16  & 15  & 11  & 5   & 1  &    &   &   &    64   \\ \hline
6  & 32  & 32  & 31  & 26  & 16  & 6  & 1  &   &   &   144   \\ \hline
7  & 64  & 64  & 63  & 57  & 42  & 22 & 7  & 1 &   &   320   \\ \hline
8  & 128 & 128 & 127 & 120 & 99  & 64 & 29 & 8 & 1 &   704   \\ 
\hline\hline
\end{tabular}
\end{center}
\label{nttable}
\end{table}

Our result (\ref{ntrr}) shows that for $d=r$, there is a unique
term $\lambda_{T,r,d=r}$.  We have 

\medskip

{\bf Corollary 2 } \quad For $G[(K_r)_m,jn]$, 
\beq
\lambda_{T,r,r} = r! \ .
\label{lamtutrr}
\eeq

\medskip

{\bf Proof} \quad  Given the uniqueness, $n_T(r,r)=1$, this follows from the
analogous term in the chromatic polynomial, eq. (\ref{lamrr}) in conjunction
with the structural equation (\ref{tgsumclan}) and the general relation 
(\ref{pt}).  $\Box$

Since $\lambda_{T,r,r}$ is unique, we omit the final index $j$, setting 
$\lambda_{T,r,r,1} \equiv \lambda_{T,r,r}$. 

The previously known calculations of $T(G[(K_r)_m,jn],x,y)$ illustrate our 
general structural theorems.   For $r=1$, the family $G[(K_1)_m,jn]$ is
identical to the circuit graph with $m$ vertices, $C_m$, and an elementary
calculation gives
\beq
T(C_m,x,y) = y + \sum_{j=1}^{m-1} x^j = \frac{1}{x-1}\left [ \mu_0 x^m + \mu_1
\right ]= \frac{1}{x-1}\left [ x^m + xy-x-y \right ] \ .
\label{tcm}
\eeq
so that $\lambda_{T,1,0}=x$ and, in accord with (\ref{lamtutrr}), 
$\lambda_{T,1,1}=1$. 

For $r=2$, the Potts model partition function and equivalent Tutte polynomial
for the family $G[(K_2)_m,jn]$ were calculated and analyzed in \cite{ka}. 
One has
\beq
T(G[(K_2)_m,jn],x,y) = \frac{1}{x-1}\biggl [ 
\sum_{j=1}^2 (\lambda_{T,2,0,j})^m + 
\mu_1 \sum_{j=1}^2 (\lambda_{T,2,1,j})^m + \mu_2 (\lambda_{T,2,2})^m \biggr ]
\label{tk2k4}
\eeq
where
\beq
\lambda_{T,2,0,j} = \frac{1}{2}\biggl [y^3+2y^2+3y+x^2+3x+2 \pm \sqrt{R_{20}} \
\biggr ] \quad {\rm for} \ j=1,2
\label{lam20j}
\eeq
\beqs
R_{20} & = & 4+12x+12y+22xy+13x^2+21y^2+6x^3+20y^3+16xy^2 \cr\cr
& & +10x^2y+x^4+10y^4-4x^2y^2-2y^3x+4y^5-2x^2y^3+y^6 
\label{rt12}
\eeqs
\beq
\lambda_{T,2,1,j} = \frac{1}{2}\biggl [ y^3+2y^2+3y+2x+4 \pm \sqrt{R_{21}} \ 
\biggr ] \quad {\rm for} \ j=1,2
\label{lam21j}
\eeq
\beq
R_{21}=16+16x+32y+4x^2+12xy+33y^2+20y^3-4xy^3+10y^4+4y^5+y^6
\label{r21}
\eeq
and $\lambda_{T,2,2}=2$, in accord with (\ref{lamtutrr}). 
Thus, $n_T(2,0)=n_T(2,1)=2$ and $n_T(2,2)=1$, as in Table \ref{nttable}. 

\medskip

Using a systematic application of the deletion-contraction property, we have
calculated the $r=3$ case, i.e., $T(G[(K_3)_m,jn),x,y)$.  Our general structure
theorem yields $n_T(3,0)=n_T(3,1)=4$, $n_T(3,2)=3$, and $n_T(3,3)=1$, and we
have 
\beqs
T(G[(K_3)_m,jn],x,y) & = & \frac{1}{x-1}\biggl [
\sum_{j=1}^4 (\lambda_{T,3,0,j})^m +
\mu_1 \sum_{j=1}^4 (\lambda_{T,3,1,j})^m \cr\cr
& + & 
\mu_2 \sum_{j=1}^3 (\lambda_{T,3,2,j})^m + \mu_3(\lambda_{T,3,3})^m \biggr ]
\label{tk3k6}
\eeqs
where 
\beq
\lambda_{T,3,0,1}=\lambda_{T,3,2,3}=3y^2(y+1)
\label{lamtut301}
\eeq
\beq
\lambda_{T,3,2,j}=\frac{3}{2}\biggl [ y^3+3y^2+2x+6y+8 \pm \sqrt{R_{32}} \ 
\biggr ] \quad {\rm for} \ j=1,2 
\label{lamtut32j}
\eeq
\beq
R_{32}=64+32x+112y+4x^2+24xy+100y^2+4xy^2+52y^3-4xy^3+21y^4+6y^5+y^6
\label{r32}
\eeq
and $\lambda_{T,3,3}=3!$ in accord with (\ref{lamtutrr}).  The other
$\lambda_{T,3,d,j}$'s are the solutions of quartic and cubic equations that are
somewhat lengthy and hence are given in the appendix. 

  In our previous works presenting other calculations of Tutte polynomials for
recursive families of graphs \cite{a,ta,hca,ka,s3a} we have given detailed
plots of the continuous accumulation set ${\cal B}$ in the $q$ plane for the $m
\to \infty$ limit of the Potts model partition function, for various values of
$v$ and ${\cal B}$ in the $v$ (or equivalent $a$) plane for various values of
$q$.  Here we shall restrict ourselves to one interesting comparison. 

We begin by working out the locus ${\cal B}_q$ for the $v=-1$ case where the
Potts model partition function reduces to the chromatic polynomial.  Here,
using (\ref{pclan}) and (\ref{lamchrom}), we find for the $m \to \infty$ limit
of the general $G[(K_r)_m,jn]$ clan graph that this continuous accumulation set
is the union of $r$ circles ${\cal C}_{r,j}$, $j=1,...,r$,
\beq 
{\cal B} = \ \quad \bigcup_{j=1}^r \ {\cal C}_{r,j}
\label{b}
\eeq
where ${\cal C}_{r,j}$ is the solution of the equation $|q-r-j+1|=r-j+1$, viz.,
\beq
{\cal C}_{r,j} \ : \quad q=r+j-1 + (r-j+1)e^{i\theta} \ , \quad
0 \le \theta < 2\pi \ , \quad 1 \le j \le r
\label{c_ell}
\eeq
The $j$'th circle ${\cal C}_{r,j}$ intersects the real $q$ axis at the points
$q=2(j-1)$ and
\beq
q=q_c(\{(K_r,K_{2r})\})=2r
\label{qc}
\eeq
Thus, these $r$ circles osculate, i.e., coincide with equal (vertical)
tangents, at $q=q_c$, which is thus a tacnodal multiple point, in the
terminology of algebraic geometry. 

The circles comprising ${\cal B}$ separate the $q$ plane
into the following $r+1$ regions $R_{r,j}$ (where ${\rm ext}({\cal C}_{r,j})$ 
and ${\rm int}({\cal C}_{r,j})$ denote the exterior and interior of the circle
${\cal C}_{r,j}$:
\beq
R_{r,1} = {\rm ext}({\cal C}_{r,1}) \ ,
\label{r1}
\eeq
\beq
R_{r,j} = {\rm int}({\cal C}_{r,j-1}) \ \bigcap \ {\rm ext}({\cal C}_{r,j}) 
\quad {\rm for} \quad 2 \le j \le r \ ,
\label{rj}
\eeq
and
\beq
R_{r,r+1} = {\rm int}({\cal C}_{r,r}) \ .
\label{rjp1}
\eeq

In region $R_{r,1}$, the $W$ function is
\beq
W = (\lambda_{P,r,0})^{1/r} \quad {\rm for} \quad q \in R_{r,1}
\label{wr1}
\eeq
In other regions, only the magnitude $|W|$ can be determined unambiguously, and
we find
\beq
|W| = |\lambda_{P,r,j-1}|^{1/r} \quad {\rm for} \quad q \in R_{r,j} \ , 
2 \le j \le r+1 \ .
\label{wrj}
\eeq

\begin{figure}[hbtp]
\centering
\leavevmode
\epsfxsize=4.0in
\begin{center}
\leavevmode
\epsffile{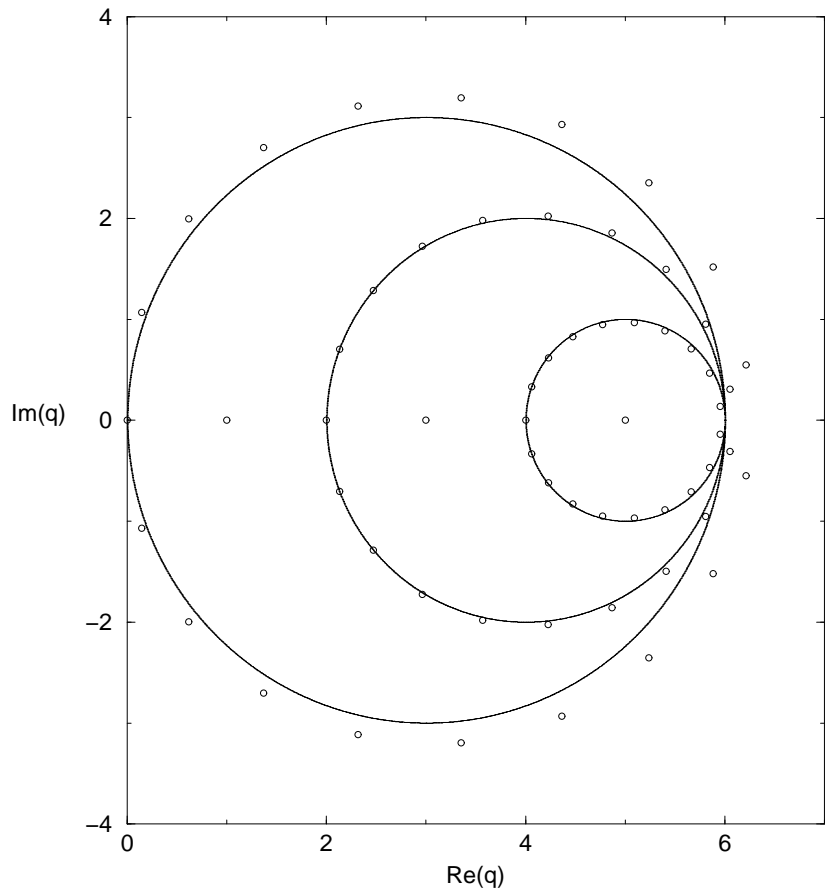}
\end{center}
\caption{\footnotesize{Singular locus ${\cal B}$ in the $q$ plane for the $m
\to \infty$ limit of the Potts model partition function $Z(G[(K_3)_m,jn],q,v)$
for $v=-1$, where this becomes the chromatic polynomial.  For comparison,
chromatic zeros for $m=20$ (i.e., $|V|=60$) are also shown.}}
\label{k3k6}
\end{figure}

We now specialize to the family $G[(K_3)_m,jn]$.  For this case, if $v=-1$, the
locus ${\cal B}_q$ consists of the union of the three circles ${\cal C}_{3,1}:
\ |q-3|=3$, ${\cal C}_{3,2}: \ |q-4|=2$, and ${\cal C}_{3,3}: \ |q-5|=1$ which
osculate at $q_c=6$.  In Fig. \ref{k3k6} we show this locus, together with
chromatic zeros calculated for the length $m=20$.  For this length, the zeros
(except for the discrete real zeros at $q=1$, $q=3$, and $q=5$) lie reasonably
close to the asymptotic curves comprising the locus ${\cal B}$.

\begin{figure}[hbtp]
\centering
\leavevmode
\epsfxsize=4.0in
\begin{center}
\leavevmode
\epsffile{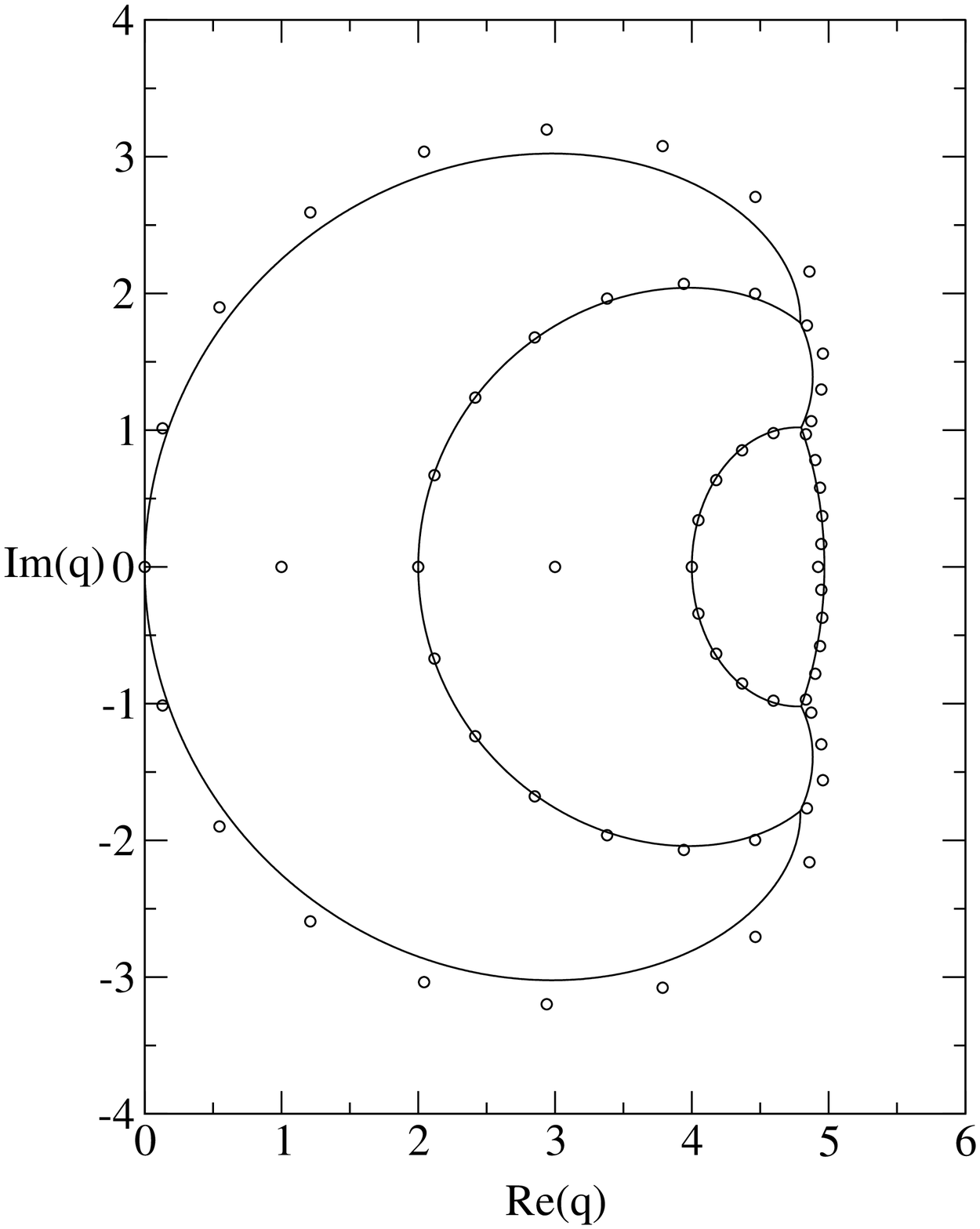}
\end{center}
\caption{\footnotesize{Singular locus ${\cal B}$ in the $q$ plane for the $m
\to \infty$ limit of the Potts model partition function $Z(G[(K_3)_m,jn],q,v)$
for $v=-0.9$. For comparison, zeros of this partition function are shown for 
$m=20$ (i.e., $|V|=60$).}}
\label{k4pxpy3a0p1}
\end{figure}

Using our new calculation of $T(G[(K_3)_m,jn],x,y)$ and the equivalent
$Z(G[(K_3)_m,jn],q,v)$ we next show in Fig. \ref{k4pxpy3a0p1} the locus ${\cal
B}_q$ for the illustrative value $v=-0.9$, together with partition function
zeros calculated for the length $m=20$.  Again, except for the real zeros at
$q=1$ and $q=3$, the zeros lie reasonably close to ${\cal B}$ for this value of
$m$. Several interesting differences are evident: (i) increasing $v$ from $-1$
to $v=-0.9$ removes the osculation point at $q=6$ and replaces it with two
complex-conjugate pairs of T-intersection points (such points were previously
encountered in many loci ${\cal B}$, e.g. \cite{z6}); (ii) the right-most
portion of ${\cal B}_q$ moves leftward, from $q=6$ at $v=-1$ to $q \simeq 5$
for $v=-0.9$; and (iii) the left-hand portion of ${\cal B}$ passes through
$q=0$ in both cases.  As with $W(\{G\},q)$ in the $v=-1$ case, different
analytic forms for the free energy apply in the different regions bounded by
the portions of ${\cal B}$ for general $v$.

\section{Flow Polynomials for $G[(K_r)_m,jn]$}

We first note the 

\medskip

{\bf Corollary 3}  
\beq
F(G[(K_r)_m,jn],q)=(-1)^{3mr(r-1)/2} \sum_{d=0}^r \mu_d \sum_{j=1}^{n_T(r,d)} 
(\lambda_{T,r,d,j}(x=0,y=1-q))^m \ .
\label{fgsumclan}
\eeq

\medskip

{\bf Proof} \quad  This follows immediately from combining eqs. (\ref{ft}) and 
(\ref{tgsumclan}) and using (\ref{ve}).  $\Box$ 

\medskip

It will be convenient to write (\ref{fgsumclan}) simply as
\beq
F(G[(K_r)_m,jn],q) = \sum_{d=0}^r \mu_d \sum_{j=1}^{n_T(r,d)} 
(\lambda_{F,r,d,j})^m
\label{flowkrgen}
\eeq
where
\beq
\lambda_{F,r,d,j}=(-1)^{3r(r-1)/2}\lambda_{T,r,d,j} \ .
\label{flamrel}
\eeq

Since $G[(K_1)_m,jn]=C_m$, it follows that $F(G[(K_1)_m,jn],q)=F(C_m,q)$.  This
is elementary: 
\beq
F(C_m,q)=q-1 \ . 
\label{flowcm}
\eeq

The flow polynomial for $G[(K_2)_m,jn]$ can be obtained from our previous
calculation of the Tutte polynomial for this family \cite{ka}:
\beq
F(G[(K_2)_m,jn],q)=\sum_{d=0}^2 \mu_d \sum_{j=1}^{n_T(2,d)}
(\lambda_{F,2,d,j})^m 
\label{flowk2}
\eeq
where $n_T(2,0)=2$, $n_T(2,1)=2$, $n_T(2,2)=1$, and 
\beq
\lambda_{F,2,0,j}=\frac{1}{2}\left [ (q-2)(q^2-3q+4) \pm \sqrt{R_{F20}} \ 
\right ] \quad {\rm for} \ \ j=1,2
\label{flam20j}
\eeq
with
\beq
R_{F20}=(q-2)(q-3)(q^4-5q^3+14q^2-20q+12)
\label{rf2}
\eeq
\beq
\lambda_{F,2,1,j}=\frac{1}{2}\left [ q^3-5q^2+10q-10
\pm \sqrt{R_{F21}} \ \right ] \quad {\rm for} \ \ j=1,2
\label{flam21j}
\eeq
where
\beq
R_{F21}=q^6-10q^5+45q^4-120q^3+208q^2-224q+116 \ .
\label{rf21}
\eeq
and
\beq
\lambda_{F,2,2,1} \equiv \lambda_{F,2,2}=-2
\label{flam22}
\eeq
The flow polynomials for the first few values of $m$ are
\beq
F(G[(K_2)_1,jn],q)=(q-1)^3(q-2)
\label{flowk2m1}
\eeq

\beq
F(G[(K_2)_2,jn],q)=(q-1)(q-2)^2(q^4-5q^3+12q^2-16q+10)
\label{flowk2m2}
\eeq

\beq
F(G[(K_2)_3,jn],q)=(q-1)(q-2)^2(q^2-4q+5)(q^5-6q^4+18q^3-34q^2+37q-28)
\label{flowk2m3}
\eeq
It is elementary to show that in general, $F(G[(K_2)_m,jn],q)$ has the factor
$(q-1)(q-2)$.

\begin{figure}[hbtp]
\centering
\leavevmode
\epsfxsize=4.0in
\begin{center}
\leavevmode
\epsffile{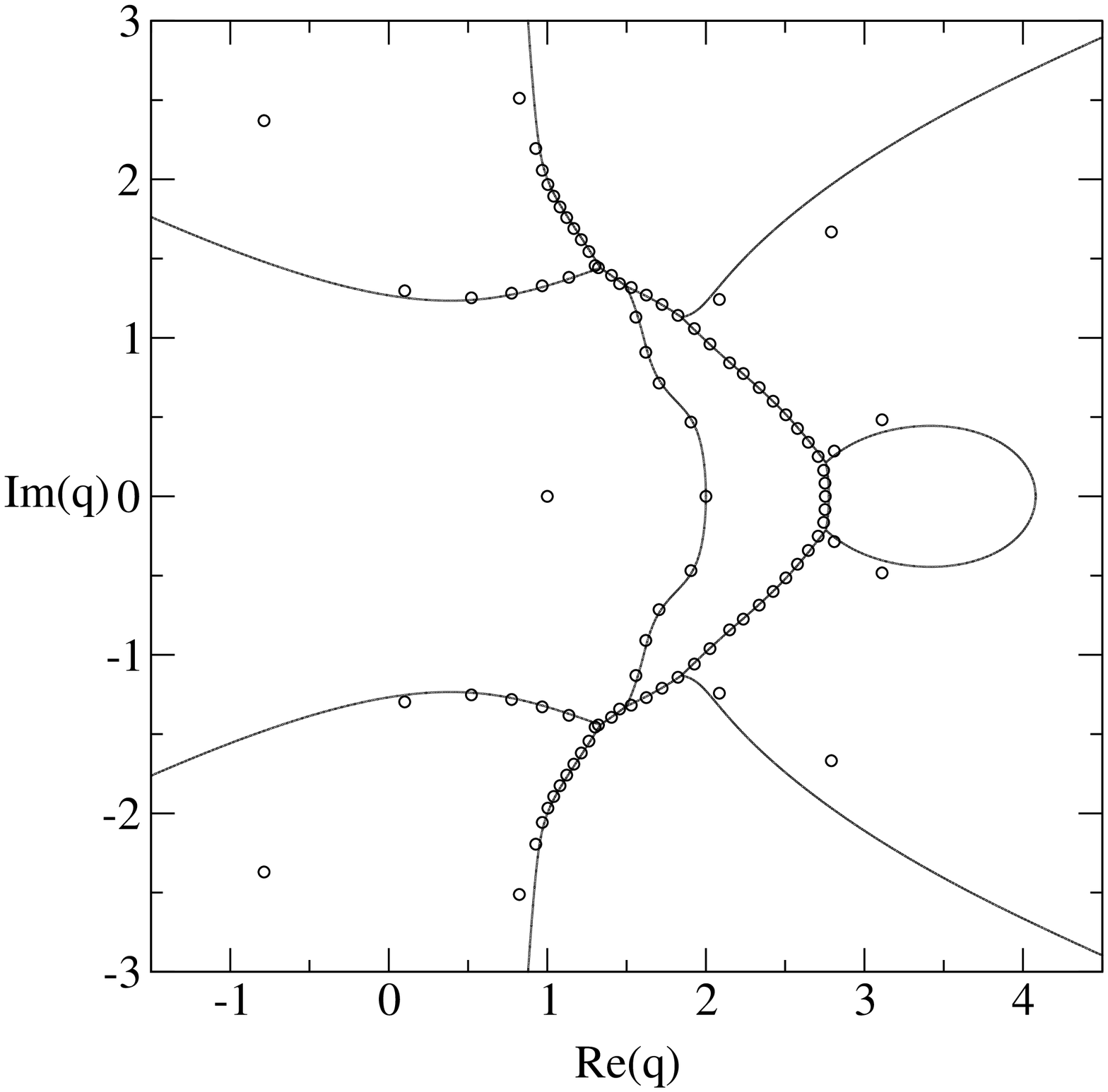}
\end{center}
\caption{\footnotesize{Singular locus ${\cal B}$ in the $q$ plane for the $m
\to \infty$ limit of the flow polynomial $F(G[(K_2)_m,jn],q)$. For comparison,
zeros of the flow polynomial $F(G[(K_2)_m,jn],q)$ are shown for $m=30$ (i.e.,
$|V|=60$).}}
\label{k4pxy2flow}
\end{figure}

In Fig. \ref{k4pxy2flow} we show the continuous accumulation set ${\cal B}$ of
the zeros of $F(G[(K_2)_m,jn],q)$ as $m \to \infty$.  Some curves on ${\cal B}$
extend to complex infinity, so that ${\cal B}$ is noncompact, in the $q$ plane.
Hence, it is also convenient to display ${\cal B}$ in the plane of the variable
\beq
u=\frac{1}{q} \ . 
\label{u}
\eeq 
The locus ${\cal B}$ separates the $q$ plane (or equivalently, the $1/q$ plane)
into several regions, including four regions that contain intervals of the real
axis, together with two pairs of complex-conjugate regions. Six curves forming
three complex-conjugate pairs lying on the locus ${\cal B}$ extend infinitely
far from the origin of the $q$ plane.  In the $u$ plane, these curves (taken in
their totality) have a multiple intersection point, in the sense of algebraic
geometry, at the origin. We introduce polar coordinates, letting
\beq
u = |u| e^{i\theta_u} \ . 
\label{upolar}
\eeq
We have

\medskip

{\bf Corollary 4 } 

\beq
\theta_u = \frac{(2j+1)\pi}{6} \quad {\rm for} \ \ 0 \le j \le 5
\label{thetau}
\eeq
i.e., $\theta_u = \pm \pi/6$, $\pm \pi/2$, and $\pm 5\pi/6$. 

\medskip

{\bf Proof} \quad To prove this, we note first that the two terms that are
dominant in the vicinity of $u=0$ are $\lambda_{F,2,0,1}$ and 
$\lambda_{F,2,1,1}$.  We extract a factor of $q^3$ from these, defining 
$\bar\lambda_{F,2,d,j} = q^{-3}\lambda_{F,2,d,j}$, express the 
$\bar\lambda_{F,2,d,j}$ in terms of $u$, and carry out a Taylor series 
expansion of these in the vicinity of $u=0$, to find
\beq
\bar\lambda_{F,2,0,1} = 1-5u+10u^2-9u^3 + O(u^4) \quad {\rm as} \ \ u \to 0
\label{lam201utaylor}
\eeq
\beq
\bar\lambda_{F,2,1,1} = 1-5u+10u^2-10u^3 + O(u^4) \quad {\rm as} \ \ u \to 0
\label{lam211utaylor}
\eeq
The equation of the degeneracy of magnitudes of leading terms in the vicinity
of $u=0$, $|\lambda_{F,2,0,1}|=|\lambda_{F,2,1,1}|$, yields the condition 
\beq
|u|^3 \cos 3\theta_u = 0 \quad {\rm for} \ \ |u| << 1
\label{cos3theq}
\eeq
implying that $\cos 3\theta_u=0$, which yields the result
(\ref{thetau}). $\Box$. 

\begin{figure}[hbtp]
\centering
\leavevmode
\epsfxsize=4.0in
\begin{center}
\leavevmode
\epsffile{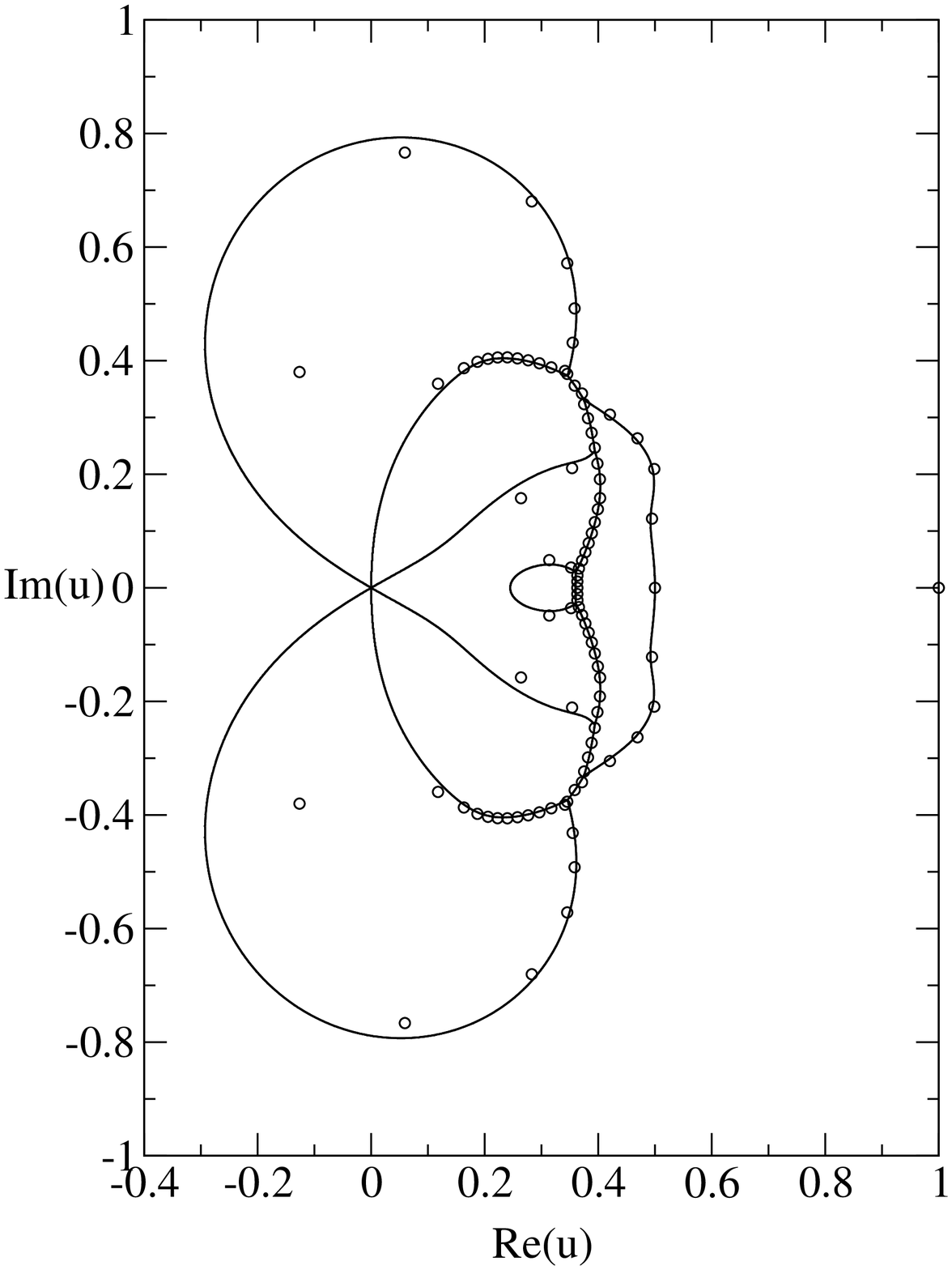}
\end{center}
\caption{\footnotesize{Singular locus ${\cal B}$ in the $u=1/q$ plane for the
$m \to \infty$ limit of the flow polynomial $F(G[(K_2)_m,jn],q)$. For
comparison, zeros of the flow polynomial $F(G[(K_2)_m,jn],q)$ are shown for
$m=30$ (i.e., $|V|=60$).}}
\label{k4pxy2uflow}
\end{figure}

The locus ${\cal B}$ crosses the real axis at three points.  The largest of
these we denote $q_c$; this has the value 
\beq
q_c(F(G[(K_2)_\infty,jn]) = 4.079828...
\label{qck2k4}
\eeq
The other two crossings occur at 
\beq
q_1=2.7760454...
\label{q1k2k4}
\eeq
and
\beq
q=2 \ .
\label{q2k2k4}
\eeq
The point $q_c$ is the larger of the two real solutions of the equation
\beq
q^4-7q^3+13q^2-q-14=0
\label{qceqk2k4}
\eeq
which is the equation of degeneracy in magnitudes of dominant
terms $|\lambda_{F,2,0,1}|=|\lambda_{F,2,1,1}|$.  The point $q_1$ is the
real solution of the equation of degeneracy in magnitudes of dominant terms 
$|\lambda_{F,2,1,1}|=|\lambda_{F,2,2}|$, 
\beq
q^3-4q^2+7q-10=0 \ .
\label{q1eqk2k4}
\eeq
The four regions that contain intervals of the real axis, together with the
respective intervals, are: (i) $R_1$, $q \ge q_c$, (ii) $R_2$, $q_1 \le q \le
q_c$, (iii) $R_3$, $2 \le q \le q_1$, and (iv) $R_4$, $q \le 2$. The two
complex-conjugate regions are denoted $R_5,R_5^*$ and $R_6,R_6^*$; in
Fig. \ref{k4pxy2flow}, as one moves counterclockwise from $R_1$ into the upper
half $q$-plane, one traverses $R_5$, containing the point $q=2+2i$, and then
$R_6$, containing $q=2i$.  As is evident in Fig. \ref{k4pxy2flow}, the density
of zeros is rather different along different portions of ${\cal B}$. 

In region $R_1$, with an appropriate choice of branch cuts, the dominant
$\lambda$ is $\lambda_{F,2,0,1}$ so that
\beq
\phi(G[(K_2)_\infty,jn],q) = (\lambda_{F,2,0,1})^{1/2} \quad {\rm for} \ \ 
q \in R_1 \ .
\label{phi_k2k4_r1}
\eeq
In region $R_2$, the dominant $\lambda$ is $\lambda_{F,2,1,1}$ so that
\beq
|\phi(G[(K_2)_\infty,jn],q)| = |\lambda_{F,2,1,1}|^{1/2} \quad {\rm for} \ \
q \in R_2 \ .
\label{phi_k2k4_r2}
\eeq
(Recall that in regions other than $R_1$, only $|\phi|$ can be obtained
unambiguously.) In region $R_3$, $\lambda_{F,2,2}$ is dominant and
\beq
|\phi(G[(K_2)_\infty,jn],q)| = \sqrt{2} \quad {\rm for} \ \ q \in R_3 \ .
\label{phi_k2k4_r3}
\eeq
In region $R_4$, with an appropriate choice of square root branch cut, 
$\lambda_{F,2,1,2}$ is dominant and 
\beq
|\phi(G[(K_2)_\infty,jn],q)| = |\lambda_{F,2,1,2}|^{1/2} \quad {\rm for} \ \
q \in R_4 \ .
\label{phi_k2k4_r4}
\eeq
In $R_5$ and $R_6$, $\lambda_{F,2,1,2}$ and $\lambda_{F,2,0,1}$ dominate,
respectively.  We also note that there are several $T$ intersection points on
${\cal B}$.  

 From our calculation of $T(G[(K_3)_m,jn],x,y)$ presented here, one can obtain
the corresponding flow polynomial $F(G[(K_3)_m,jn],q)$.  Since the expressions
are rather lengthy, we omit the details.

\section{Reliability Polynomials for $G[(K_r)_m,jn]$} 

We proceed to discuss reliability polynomials for $G[(K_r)_m,jn]$. 
Since $G[(K_1)_m,jn]$ is the circuit graph $C_m$, the result is elementary:
\beq
R(C_m,p) = p^{m-1}[m(1-p)+p] \ .
\label{rcn}
\eeq
Note that when one sets $x=1$, the two $\lambda_{T,r=1,d,j}$ simplify:
$\lambda_{T,1,0,1}=x$ becomes equal to $\lambda_{T,1,1,1}=1$, so that Hence,
the total number of distinct terms $N_{T,r=1,\lambda}=2$ in the Tutte
polynomial is reduced to $N_{R,r=1,\lambda}=1$ for the reliability polynomial.
This reduction generalizes to higher $r$.  The reliability polynomial
$R(C_m,p)$ evidently has $m-1$ zeros at $p=0$ and one other zero, at
$p=m/(m-1)$.  As $m \to \infty$, this other zero approaches 1 from above.

The reliability polynomial for $G[(K_2)_m,jn]$ can be obtained from our
previous calculation of the Tutte polynomial for this family of graphs
\cite{ka}.  Setting $x=1$, we find that two of the five $\lambda$'s in 
the full Tutte polynomial become equal to two others: 
\beq
\lambda_{T,2,0,1}=\lambda_{T,2,1,1}=\frac{1}{2}\left [(y+2)(y^2+3)+\sqrt{R_2}
 \ \right ] 
\label{lam201x1}
\eeq
\beq
\lambda_{T,2,0,2}=\lambda_{T,2,1,2}=\frac{1}{2}\left [(y+2)(y^2+3)-\sqrt{R_2}
 \ \right ]
\label{lam202x1}
\eeq
where
\beq
R_2 = 36+44y+33y^2+16y^3+10y^4+4y^5+y^6 \ .
\label{r2pol}
\eeq
Hence, $N_{T,2,\lambda}=5$ is reduced to $N_{R,2,\lambda}=3$ for the
reliability polynomial. Setting $y=1/(1-p)$ and inserting the appropriate
prefactors according to (\ref{reliability}), we obtain
\beqs
& & R(G[(K_2)_m,jn],p) = p^{2m}\biggl [ (\alpha_{2,1})^m + (\alpha_{2,2})^m 
-\frac{3}{2}(\alpha_{2,3})^m + \cr\cr
& & mp^{-1}(1-p)^4 \biggl \{ (\alpha_{2,1})^{m-1} D_{2,1} + 
                             (\alpha_{2,2})^{m-1} D_{2,2} \biggr \} \biggr ]
\label{rk2}
\eeqs
where
\beq
\alpha_{2,j} = \frac{1}{2}\Bigl [ (3-2p)(4-6p+3p^2) \pm \sqrt{R_{2p}}\ \Bigr ] 
\quad {\rm for} \ \ j=1,2
\label{alpha2j}
\eeq
\beq
R_{2p}=36p^6-260p^5+793p^4-1308p^3+1236p^2-640p+144
\label{r2}
\eeq
\beq
D_{2,j} = \frac{1}{2}\left [ 3 \pm \frac{(36-88p+69p^2-18p^3)}
{\sqrt{R_{2p}}} \ \right ] \quad {\rm for} \ \ j=1,2 
\label{d2j}
\eeq
\beq
\alpha_{2,3}=2(1-p)^3 \ .
\label{alpha23}
\eeq
As examples, skipping the degenerate case $m=1$, the reliability polynomials
for $G[(K_2)_2,jn]$ for the lowest two cases $m=2,3$ are
\beq
R(G[(K_2)_2,jn],p)=p^3(2-p)(6p^6-44p^5+139p^4-242p^3+246p^2-140p+36)
\label{rk2m2}
\eeq
and
\beqs
& &R(G[(K_2)_3,jn],p)=p^5(120p^{10}-1440p^9+7830p^8-25440p^7+54780p^6-81840p^5 
\cr\cr
& & +86110p^4-63195p^3+31080p^2-9300p+1296)
\label{rk2m3}
\eeqs

\begin{figure}[hbtp]
\centering
\leavevmode
\epsfxsize=4.0in
\begin{center}
\leavevmode
\epsffile{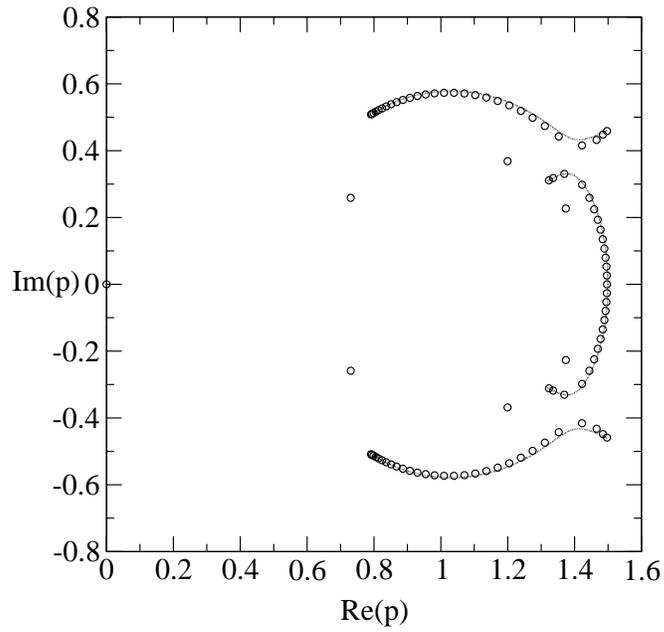}
\end{center}
\caption{\footnotesize{Singular locus ${\cal B}$ in the complex $p$ plane for
the $m \to \infty$ limit of the reliability polynomial $R(G[(K_2)_m,jn],p)$. 
For comparison, zeros of the reliability polynomial $R(G[(K_2)_m,jn],p)$ are 
shown for $m=30$ (i.e., $|V|=60$).}}
\label{k4pxy2rel}
\end{figure}

In Fig. \ref{k4pxy2rel} we show the continuous accumulation set ${\cal B}$ of
the zeros of $R(G[(K_2)_m,jn],p)$ in the limit $m \to \infty$.  This locus is
comprised of a self-conjugate arc that crosses the real axis at $p=3/2$ and
terminates at approximately $p = 1.320 \pm 0.309i$, together with a
complex-conjugate pair of arcs.  The upper arc extends between endpoints at
$0.790 + 0.508i$ and $1.501 + 0.462i$.  The six endpoints of the arcs occur at
the zeros of $R_{2p}$, where there are square root branch point singularities
in some of the $\lambda$'s.  This locus ${\cal B}$ does not separate the $p$
plane into different regions.  In passing, we observe that our results are in
accord with the conjecture \cite{bc} that for an arbitrary (connected) graph
$G$, the roots of $R(G,p)$ lie in the disk $|p-1| \le 1$.

\vspace{16mm}

\section{Acyclic Orientations for Recursive Families of Graphs}

\subsection{General} 

We recall from the introduction the definition of acyclic orientations and the
relations (\ref{ap}) and (\ref{at}) expressing the number of acyclic
orientations $a(G)$ in terms of a valuation of the chromatic or Tutte
polynomial.  First we note that for the recursive families of graphs of
interest here, namely cyclic or cylindrical strips of regular lattices, 
$a(G_m)$ grows exponentially as $m \to \infty$. 
This follows from the relation (\ref{ap}) and the structural
result (\ref{pgsum}).  This motivates the definition 

\medskip

{\bf Def.} \quad Consider a recursive family of graphs $G_m$ and the limit $m
\to \infty$.  Then we define 
\beq
\alpha(\{G\}) = \lim_{|V| \to \infty} a(G)^{1/|V|} \ .
\label{alpha}
\eeq

\medskip

The number of acyclic orientations can also be generalized \cite{stanley} 
to $a_s(G)$, given by 
\beq
a_s(G)=(-1)^{|V|}P(G,q=-s) \quad {\rm for} \ \ s \in {\mathbb Z}_+ \ . 
\label{asgp}
\eeq
with $a_1(G) \equiv a(G)$. 

{\bf Def.} Let $G_m$ be a recursive family of graphs, such as strips of regular
lattices of length $m$, and consider the limit $m \to \infty$.  Then 
\beq
\alpha_s(\{G\}) = \lim_{|V| \to \infty}(a_s(G))^{1/|V|} \ .
\label{alpham}
\eeq
Hence, 
\beq
\alpha_s(\{G\})=W(\{G\},q=-s) \ .
\label{alphamw}
\eeq
and, in particular, $\alpha(\{G\})=W(\{G\},q=-1)$. 

In our previous studies of chromatic polynomials and the resultant singular
loci ${\cal B}$ for a variety of recursive families of graphs including strips
of regular lattices, we found that ${\cal B}$ never intersects the negative
real $q$ axis, and, indeed, the points $q=-s$ for $s \in {\mathbb Z}_+$ are in
region $R_1$.  Hence, for the recursive families of interest here, there is no
ambiguity in which $1/|V|$'th root to choose in evaluating (\ref{w}) at the
points $q=-s$, $s \in {\mathbb Z}_+$.  (For families with noncompact ${\cal
B}$, as is evident from the plots of respective loci ${\cal B}$ presented in
\cite{w,wa,wa2,wa3}, although ${\cal B}$ does not intersect the negative real
axis, it can separate the point $q=-s$ from the region $R_1$; we do not
consider such families here.)  Because the points $q < 0$ are in region $R_1$
for the strip graphs of regular lattices, we can apply a previous result
\cite{bcc} to infer that $\alpha_s(\{G\})$ is independent of the longitudinal
boundary conditions.  Thus, for example, $\alpha_s(\{G\})$ is the same for the
$m \to \infty$ limit of the square-lattice strip graphs
$sq(L_y,L_x,FBC_y,FBC_x)$, $sq(L_y,L_x,FBC_y,PBC_x)$, and
$sq(L_y,L_x,FBC_y,TPBC_x)$, and therefore we shall omit the longitudinal
boundary condition in listing the values that we obtain in these cases.
Henceforth, we shall focus on the $s=1$ case, i.e., acyclic orientations.

\subsection{$a(G)$ and $\alpha(\{G\})$ for Strips of the Square Lattice}

We first list some elementary results. For a tree graph, 
\beq
a(T_n)=2^{n-1} \ .
\label{atn}
\eeq
For a circuit graph,
\beq
a(C_n)=2^n-2
\label{acn}
\eeq
whence
\beq
\alpha(\{T\})=\alpha(\{C\})=\alpha(sq,L_y=1)=2 \ .
\label{alphasqly1f}
\eeq
Another easy result is 
\beq
a(sq,L_y=2,L_x=m,FBC_y,FBC_x)=2 \cdot 7^m
\label{asqly2f}
\eeq
 From $P(sq[L_y=2,L_x=m,FBC_y,(T)PBC_x],q)$ calculated in \cite{bds}, we
find
\beq
a(sq,L_y=2,L_x=m,FBC_y,PBC_x)=7^m-2 \cdot 4^m-2^{m+1}+5
\label{asqly2p}
\eeq
\beq
a(sq,L_y=2,L_x=m,FBC_y,TPBC_x)=7^m-2 \cdot 4^m + 2^{m+1}-1
\label{asqly2mb}
\eeq
 From any of (\ref{asqly2f}), (\ref{asqly2p}), or (\ref{asqly2mb}), it follows
 that 
\beq
\alpha(sq,L_y=2,FBC_y)=\sqrt{7}=2.645751...
\label{alphasqly2f}
\eeq
From $P(sq[L_y=3,L_x=m,FBC_y,PBC_x],q)$ calculated in \cite{wcyl,wcy}, we 
find
\beqs
a(sq,L_y=3,L_x=m,FBC_y,PBC_x) & = & -13+5(2^m+3^m+5^m)-2 \cdot 9^m -2S_m
\cr\cr & & + \Biggl ( \frac{27+\sqrt{481}}{2} \Biggr )^m + \Biggl ( 
\frac{27-\sqrt{481}}{2} \Biggr )^m
\label{asqly3cyc}
\eeqs
and from $P(sq[L_y=3,L_x=m,FBC_y,TPBC_x],q)$ calculated in \cite{pm}, we 
find
\beqs
a(sq,L_y=3,L_x=m,FBC_y,TPBC_x) & = & 5-2^m+3^m-5^m+2 \cdot 9^m -2S_m
\cr\cr & &+ \Biggl ( \frac{27+\sqrt{481}}{2} \Biggr )^m + \Biggl (
\frac{27-\sqrt{481}}{2} \Biggr )^m
\label{asqly3mb}
\eeqs
with 
\beq
S_m = \sum_{j=1}^3 (-\lambda_{sqpxy3,j})^m
\label{sm}
\eeq
where the $\lambda_{sqpxy3,j}$, $j=1,2,3$, are the roots of the equation
\beq
\xi^3+23 \xi^2+134 \xi+202=0
\label{eqcubsqcyc}
\eeq 
 From either of (\ref{asqly3cyc}) or (\ref{asqly3mb}) we have 
\beq
\alpha(sq,L_y=3,FBC_y)=\Biggl ( \frac{27+\sqrt{481} \ }{2}
\Biggr )^{1/3} = 2.903043...
\label{alphasqly3f}
\eeq

 From the chromatic polynomial $P(sq[L_y=4,L_x,FBC_y,FBC_x],q)$ calculated in
\cite{strip}, we obtain 
\beq
\alpha(sq,L_y=4,FBC_y)=(\lambda_{sqxy4,max})^{1/4}=3.040731...
\label{alphasqly4f}
\eeq
where $\lambda_{sqxy4,max}$ is the largest root of the cubic equation
\beq
\xi^3-105\xi^2+1747\xi-6758=0 \ . 
\label{sqxy4cubeq}
\eeq 
 From chromatic polynomials calculated in \cite{s4} and \cite{ss00}, we
obtain the values of $\alpha(sq,L_y,FBC_y)$ for $L_y$ up to 8. The results are
listed in Table \ref{alphasq}. 

We next consider strips of the square lattice with cylindrical boundary 
conditions.  An elementary calculation yields 
\beq
a(sq,L_y=3,L_x=m,PBC_y,FBC_x)=6 \cdot (34)^m
\label{asqly3cyl}
\eeq
whence
\beq
\alpha(sq,L_y=3,PBC_y)=(34)^{1/3}=3.2396118...
\label{alphasqly3cyl}
\eeq
 From the chromatic polynomial $P(sq[L_y=4,L_x,PBC_y,FBC_x],q)$
calculated in \cite{strip2} we obtain 
\beq
\alpha(sq,L_y=4,PBC_y)=\Biggl ( \frac{139 + \sqrt{16009} \ }{2} \Biggr
)^{1/4} 
= 3.3944509...
\label{alphasqly4cyl}
\eeq
 From the chromatic polynomials $P(sq[L_y,L_x,PBC_y,FBC_x],q)$ for $L_y=5,6$ 
calculated in \cite{strip2} we compute 
\beq
\alpha(sq,L_y=5,PBC_y)=\Biggl ( \frac{527 + \sqrt{200585} \ }{2} \Biggr
)^{1/5}=3.4481257...
\label{alphasqly5cyl}
\eeq
and
\beq
\alpha(sq,L_y=6,PBC_y)=(\lambda_{sqpxpy6,max})^{1/6}=3.4705457...
\label{alphasqly6cyl}
\eeq
where $\lambda_{sqpxpy6,max}$ is the largest root of the equation
\beq
\xi^5-2049\xi^4+547805\xi^3-36633324\xi^2+639262524\xi-2756653440=0 \ .
\label{eqsqxpy6qm1}
\eeq
We have also computed $\alpha(sq,L_y,PBC_y)$ for $L_y$ up to 12 using 
chromatic polynomials calculated in \cite{sqcyl}.  The results are listed 
in Table \ref{alphasq}. 

\begin{table}
\caption{\footnotesize{Values of $\alpha(\{G\})$ for 
the infinite-length limits of strip graphs of the square lattice with width
$L_y$ vertices and free (F) or periodic (P) transverse boundary conditions.}}
\begin{center}
\begin{tabular}{|c|c|c|}
\hline\hline
$L_y$ & $BC_y$ & $\alpha(\{G\})$ \\
\hline\hline
1  & F & 2        \\ \hline
2  & F & 2.646    \\ \hline
3  & F & 2.903    \\ \hline
4  & F & 3.041    \\ \hline
5  & F & 3.126    \\ \hline
6  & F & 3.185    \\ \hline
7  & F & 3.227    \\ \hline
8  & F & 3.259    \\ \hline\hline
3  & P & 3.240    \\ \hline
4  & P & 3.394    \\ \hline
5  & P & 3.448    \\ \hline
6  & P & 3.471    \\ \hline
7  & P & 3.481    \\ \hline
8  & P & 3.487    \\ \hline
9  & P & 3.490    \\ \hline
10 & P & 3.491    \\ \hline
11 & P & 3.492    \\ \hline
12 & P & 3.493    \\ \hline\hline 
\end{tabular}
\end{center}
\label{alphasq}
\end{table}

A basic property of $\alpha(sq,L_y,BC_y)$, where $BC_y$ denotes the transverse
boundary condition, (free, $FBC_y$ or periodic, $PBC_y$) is that, owing to the
fact that one can calculate this quantity (and indeed, more generally, the
chromatic or Tutte polynomial from which it is derived by a special valuation)
by transfer matrix methods, it follows that $\alpha(sq,L_y,BC_y)$ is a
monotonically increasing function of $L_y$ \cite{mw,cmnn}.  A similar
monotonicity holds for $\alpha(tri,L_y,BC_y)$ for strips of the triangular
lattice (see further below).  We also observe that for a given strip width
$L_y$, the value of $\alpha$ for periodic transverse boundary conditions is
greater than the value for free transverse boundary conditions.  This is easily
understood since the transverse boundary effects are reduced when one uses
periodic transverse boundary conditions, and hence the resultant $\alpha$
should be closer to that for the infinite square lattice.  Let us denote
$\alpha_{sq} \equiv \lim_{L_y \to \infty} \alpha(sq,L_y,BC_y)$.  Given the
monotonically increasing property of $\alpha(sq,L_y,BC_y)$ as a function of
$L_y$, our results show that $\alpha_{sq} > 3.49287$ (rounded off to 3.493 in
the table) for this lattice.  This value is a slight improvement of the recent
lower bounds $\alpha_{sq} \ge 22/7 = 3.1428...$ \cite{mw} and $\alpha_{sq} \ge
3.41358$ \cite{cmnn}.  Since the upper bounds $\alpha_{sq} \le 3.70925$
\cite{mw} and $\alpha_{sq} \le 3.55449$ \cite{cmnn} have also been established,
the value of $\alpha_{sq}$ has now been rather well restricted.

\subsection{$a$ and $\alpha$ for Strips of the Triangular Lattice}

For a strip of the triangular lattice, formed by starting with a square strip
and adding diagonal edges, say from the upper left to the lower right vertices
of each square, with free longitudinal boundary conditions, an elementary
calculation yields 
\beq
a(tri,L_y=2,L_x=m,FBC_y,FBC_x)=2 \cdot 3^{2m} \ .
\label{atrily2f}
\eeq
For cyclic and M\"obius strips of the triangular lattice with width $L_y=2$,
from the chromatic polynomials in \cite{wcyl}, we find 
\beq
a(tri,L_y=2,L_x=m,FBC_y,PBC_x)=3^{2m} -2 \Biggl [ 
\biggl ( \frac{7 + \sqrt{13} \ }{2} \biggr )^m + 
\biggl ( \frac{7 - \sqrt{13} \ }{2} \biggr )^m \Biggr ] + 5
\label{atrily2cyc}
\eeq
\beq
a(tri,L_y=2,L_x=m,FBC_y,TPBC_x)=3^{2m} +\frac{8}{\sqrt{13}}\Biggl [
-\biggl ( \frac{7 + \sqrt{13} \ }{2} \biggr )^m +
\biggl ( \frac{7 - \sqrt{13} \ }{2} \biggr )^m \Biggr ] -1
\label{atrily2mb}
\eeq
Using any of (\ref{atrily2f}), (\ref{atrily2cyc}), or (\ref{atrily2mb}), we
compute 
\beq
\alpha(tri,L_y=2,FBC_y)=3
\label{alphatryly2f}
\eeq
For $L_y=3$, using \cite{t}, we get 
\beqs
a(tri,L_y=3, L_x=m, FBC_y, PBC_x) & = & -13 + 5 \Biggl [ 3^m + \Biggl (
\frac{9 + \sqrt{33} \ }{2} \Biggr )^m + \Biggl ( \frac{9 - \sqrt{33} \
}{2} \Biggr )^m \Biggr ] \cr\cr
& & -2(12^m + R_m) + \Biggl ( \frac{43+\sqrt{1417} \ }{2} \Biggr )^m + 
\Biggl ( \frac{43-\sqrt{1417} \ }{2} \Biggr )^m \cr\cr
& & 
\label{atrily3cyc}
\eeqs
with 
\beq
R_m = \sum_{j=1}^3 (-\lambda_{t3c,j})^m
\label{rm}
\eeq
where the $\lambda_{t3c,j}$, $j=1,2,3$, are the roots of the equation
\beq
\xi^3+33 \xi^2+201 \xi+324=0
\label{eqcubtricyc}
\eeq
whence
\beq
\alpha(tri,L_y=3,FBC_y)=\biggl ( \frac{43+\sqrt{1417} \ }{2} \biggr
)^{1/3} = 3.4290909...
\label{alphatrixy3}
\eeq
and, for $L_y=4,5$, 
\beq
\alpha(tri,L_y=4,FBC_y)=(\lambda_{trixy4,max})^{1/4}=3.665350...
\label{alphatrixy4}
\eeq
where $\lambda_{trixy4,max}$ is the largest root of the quartic equation
\beq
\xi^4-217\xi^3+6960\xi^2-67968\xi+186624=0
\label{eqtrixy4}
\eeq
\beq
\alpha(tri,L_y=5,FBC_y)=(-\lambda_{trixy5,max})^{1/5}=3.81486986...
\label{alphatrixy5}
\eeq
where $\lambda_{txy5,max}$ is the root with the largest magnitude of the
equation
\beqs
& & \xi^9+1160\xi^8+330071\xi^7+39854484\xi^6+2509883184\xi^5
+87798785472\xi^4+1684802267136\xi^3\cr\cr
& & +16500631191552\xi^2+73085995450368\xi+104764180267008=0 \cr\cr & & 
\label{eqtrixy5}
\eeqs

For strips of the triangular lattice with cylindrical boundary conditions, 
one has 
\beq
a(tri,L_y=3,L_x=m,PBC_y,FBC_x)=6 \cdot (71)^m
\label{atrily3cyl}
\eeq
\beq
\alpha(tri,L_y=3,PBC_y)=(71)^{1/3}=4.1408177...
\label{alphatrily3cyl}
\eeq
For $L_y=4$ from the chromatic polynomial calculated in \cite{strip2} we get 
\beq
\alpha(tri,L_y=4,PBC_y)=(352)^{1/4}=4.3314735...
\label{alphatrily4cyl}
\eeq
Further, using the chromatic polynomials calculated in \cite{t}, we have 
\beq
\alpha(tri,L_y=5,PBC_y)= (897+339\sqrt{5} \ )^{1/5}=4.403125...
\label{alphatrixpy5}
\eeq
\beq 
\alpha(tri,L_y=6,PBC_y)=(\lambda_{txpy6,max})^{1/6}=4.435287...
\label{alphatrixpy6}
\eeq
where $\lambda_{txpy6,max}$ is the largest root of the equation
\beq 
\xi^5-8789\xi^4+9102104\xi^3-1119801408\xi^2+42084913152\xi-364032294912=0
\ .
\label{alphatrixpy6eq}
\eeq
The monotonicity of $\alpha(tri,L_y,FBC_y)$ and $\alpha(tri,L_y,PBC_y)$ as a
function of $L_y$ is evident in the table.  One can, of course, calculate these
quantities for larger values of $L_y$; however, it is interesting to observe
that our exact calculation of $\alpha(tri,L_y=6,PBC_y)$ is already within about
1 \% of the exact value for the $L_y \to \infty$ limit, i.e., for the infinite
triangular lattice, which we have obtained via an evaluation of the $W$
function in \cite{baxter}:
\beq
\alpha(tri)=4.474647...
\label{alphatri}
\eeq
This shows that the values of $\alpha(tri,L_y,PBC_y)$ converge rather rapidly
to the infinite-$L_y$ value as $L_y$ increases.

\begin{table}
\caption{\footnotesize{Values of $\alpha(\{G\})$ for the infinite-length 
limits of strip graphs of the triangular lattice with width
$L_y$ vertices and free (F) or periodic (P) transverse boundary conditions.}}
\begin{center}
\begin{tabular}{|c|c|c|}
\hline\hline
$L_y$ & $BC_y$ & $\alpha(\{G\}$ \\
\hline\hline
2  & F & 3       \\ \hline
3  & F & 3.429    \\ \hline
4  & F & 3.665   \\ \hline
5  & F & 3.815   \\ \hline\hline
3  & P & 4.141    \\ \hline
4  & P & 4.331   \\ \hline
5  & P & 4.403    \\ \hline
6  & P & 4.435    \\ \hline
$\infty$ 
   & P & 4.475   \\ \hline\hline
\end{tabular}
\end{center}
\label{alphat}
\end{table}

\subsection{$a$ and $\alpha$ for Cyclic Clan Graphs} 

 From eqs. (\ref{pclan}) and (\ref{lamchrom}) we have 
\beqs
a(G[(K_r)_m,jn]) & = & (-1)^{rm}P(G[(K_r)_m,jn],q=-1) \cr\cr
& = & (-1)^{rm}\left [ [(-1-r)_{(r)}]^m + 2\sum_{d=1}^{r} (-1)^d 
(\lambda_{P,r,d}(q=-1))^m \right ] \cr\cr
& = & [(2r)_{(r)}]^m + 2\sum_{d=1}^r (-1)^d [r_{(d)}(2r)_{(r-d)}]^m 
\label{acyc_clan}
\eeqs
where we have used
\beq
\lambda_{P,r,d}(q=-1) = (-1)^r r_{(d)}(2r)_{(r-d)}
\label{lamqm1}
\eeq
and
\beq
\mu_d(q=-1)=(-1)^d 2  \quad {\rm for } \ d \ge 1 
\label{mudqm1}
\eeq
(the $d=0$ coefficient being a constant, $\mu_0=1$). 

Since the point $q=-1$ is in region $R_1$, the term $\lambda_{P,r,d}$ with
$d=0$ is dominant, and hence 
\beq
\alpha(G[\{(K_r)\},jn] = [(2r)_{(r)}]^{1/r} \ .
\label{alphakr}
\eeq
For large $r \to \infty$, this has the leading asymptotic behavior 
$\alpha(G[\{(K_r)\},jn]) \sim 4e^{-1}r$. 

\section{Number of Spanning Trees for Two $G[(K_r)_m,L]$ Families}

\subsection{General} 

In this section we shall again consider only connected graphs $G=(V,E)$ and for
notational simplicity, we let 
\beq 
n=|V| \ . 
\label{nv}
\eeq
We denote the number of spanning trees of such a graph as $N_{ST}(G)$.  Here we
shall prove theorems that determine the the numbers of spanning trees of two
$G[(K_r)_m,L]$ families of graphs, namely the cyclic clan graph family
$G[(K_r)_m,L=jn]$ and the family $G[(K_r)_m,L=id]$.  In general $N_{ST}(G)$ is
given by (\ref{t11}); however, we shall actually use an alternative method of
calculation based on the Laplacian matrix (e.g., \cite{bbook}). We recall that
the adjacency matrix ${\bf A}(G)$ of a graph $G$ is the $n \times n$ matrix
whose $jk$'th element is the number of edges connecting vertex $j$ with vertex
$k$ in $V$.  Next, we recall

\medskip

{\bf Def.} \quad The Laplacian matrix ${\bf Q}(G)$ of this graph is given by 
\beq
{\bf Q}(G) = {\bf \Delta}(G) - {\bf A}(G)
\label{q}
\eeq
where ${\bf \Delta}$ is the $ n \times n$ diagonal matrix whose $j$'th
diagonal entry is equal to the degree of the $j$'th vertex, $\Delta_{jj} =
\Delta(v_j)$ and whose other entries are zero.

Since the sum of the elements in each row (or column) of ${\bf Q}(G)$ 
vanishes, one of the eigenvalues of ${\bf Q}(G)$ is zero.  Denote the
remaining $n-1$ eigenvalues by $\lambda_1,...,\lambda_{n-1}$. Then a 
basic theorem is (e.g., \cite{bbook,cvet}) 
\beqs
N_{ST}(G) & = & {\rm any \ cofactor \ of \ } {\bf Q}(G) \label{eigen} \\ 
& = & \frac{1}{n}\prod_{j=1}^{n-1}\lambda_j \ .
\label{muprod}
\eeqs
For the recursive families $G_m$ of graphs considered here, $N_{ST}(G)$ grows
exponentially for large $m$, and hence a quantity of interest is the growth
rate.  In previous works calculating $z(\{G\})$ and $N_{ST}(G)$ for the 
$n \to \infty$ limits of regular lattice graphs \cite{wu77}-\cite{sw} the
convention was used of defining this growth rate in terms of $\ln
N_{ST}^{1/n}$, and we shall follow this convention here: 

\medskip

{\bf Def.} \quad  Let $G$ be a connected graph.  Then the quantity $z(\{G\})$ 
is defined by 
\beq
\exp(z(\{G\})) = \lim_{n \to \infty} [N_{ST}(G)]^{1/n} \ . 
\label{zg}
\eeq

Since we shall compare the expressions that we derive for $N_{ST}$ on 
$G[(K_r)_m,jn]$ and $G[(K_r)_m,id]$ with upper bounds, we recall the statements
of these bounds.  A general upper bound is \cite{grim}
\beq
N_{ST}(G) \le \frac{1}{n}\biggl ( \frac{2|E|}{n-1}\biggr )^{n-1} \ . 
\label{kboundg}
\eeq
For a $\Delta$-regular graph $G$, using the relation $\Delta = 2|E|/n$, 
this implies the upper bound
\beq
N_{ST}(G) \le \frac{1}{n}\biggl ( \frac{n\Delta}{n-1} \biggr )^{n-1}
\label{kboundgregular}
\eeq
and hence
\beq
\exp(z(\{G\})) \leq \Delta \ .
\label{zgrim}
\eeq

A stronger upper bound for $\Delta$-regular graphs with vertex degree 
$\Delta \ge 3$ is \cite{mckay,cy}:
\beq
N_{ST}(G) \le \Biggl ( \frac{2\ln n}{n \Delta \ln \Delta}
\Bigg) (C_\Delta)^n,
\label{nmckay}
\eeq
where
\beq
C_\Delta = \frac{(\Delta-1)^{\Delta-1}}{[\Delta(\Delta-2)]^{\Delta/2-1}} \ .
\label{ck}
\eeq
yielding the upper bound for a $\Delta$-regular graph with $\Delta \ge 3$
\beq
\exp(z(\{G\})) \le C_\Delta \ . 
\label{zmckay}
\eeq
By expanding $C_\Delta$ as $\Delta \to \infty$, one sees that in this limit the
upper bound (\ref{zmckay}) approaches (\ref{zgrim}). 

For the comparison of a given growth rate of a $\Delta$-regular graph with the
two upper bounds (\ref{zgrim}) and (\ref{zmckay}), labelled as $u.b.,j$ with 
$j=1,2$, we define the ratios
\beq
R_j(\{G\}) = \frac{\exp(z(\{G\}))}{\exp(z(\{G\})_{u.b.,j})} \ . 
\label{rjg}
\eeq

\subsection{Number of Spanning Trees in $G[(K_r)_m,jn]$} 

Here we shall calculate the number of spanning trees for the graph 
$G[(K_r)_m,jn]$. Let ${\bf a}(j,j^\prime)$ be the $r \times r$
adjacency matrix between the vertices of $K_{r_j}$ and
$K_{r_{j^\prime}}$. The non-zero entries in this matrix are 
\beq
{\bf a}(j,j) = {\bf J}_r - {\bf I}_r \quad {\rm for} \ 1 \le j \le m 
\label{a0}
\eeq
and
\beq
{\bf a}(j,j+1) = {\bf a}(j+1,j) = {\bf J}_r \quad {\rm for} \ 1 \le j \le 
m
\label{a1}
\eeq
where ${\bf J}_r$ is the $r \times r$ matrix with all elements equal to unity, 
and ${\bf I}_r$ is the $r \times r$ identity matrix. The Laplacian matrix
of the graph $G[(K_r)_m,jn]$ is then 
\beqs
{\bf Q}(G[(K_r)_m,jn]) &=&  [ (3r-1){\bf I}_r - {\bf a}(j,j) ] \otimes
{\bf I}_m - {\bf a}(j,j+1) \otimes {\bf R}_m - {\bf a}(j+1,j) \otimes {\bf
R}^T_m \nonumber \\
&=& (3r{\bf I}_r - {\bf J}_r) \otimes {\bf I}_m - {\bf J}_r \otimes {\bf
R}_m - {\bf J}_r \otimes {\bf R}^T_m 
\label{qkb}
\eeqs
where ${\bf R}_m$ is the $m \times m$  matrix
\beq
{\bf R}_m=\left(\begin{array}{ccccc}
0&1&0&\cdots&0\\
0&0&1&\cdots&0\\
\vdots&\vdots&\vdots&\ddots&\vdots\\
0&0&0&\cdots&1\\
1&0&0&\cdots&0
\end{array}
\right) \ .
\label{rn}
\eeq

\medskip

{\bf Lemma 3} \quad The eigenvalues of ${\bf Q}(G[(K_r)_m,jn])$ are 
\beq
\lambda({\bf Q}(G[(K_r)_m,jn])) = \cases{3r - (r+re^{i2\pi j/m}+ 
re^{-i2\pi j/m}) = 2r[1-\cos(2\pi j/m)] \ {\rm for} \ j = 0,1,\cdots,m-1
\cr\cr
3r \quad \quad {\rm with \ multiplicity} \ (r-1)m.}
\label{eigenqkb}
\eeq

\medskip

{\bf Proof} \quad  We first observe that 
${\bf R}_m$  can be diagonalized by the similarity transformation  ${\bf
S}_m {\bf R}_m {\bf S}_m^{-1}$ generated by the matrix ${\bf S}_m$ with
elements
\beq
({\bf S}_m)_{jk} = ({\bf S}_m^{-1})^*_{jk} = m^{-1/2}e^{i2\pi jk/m} \quad
{\rm for} \ j,k=0,1,\cdots, m-1,
\label{sn}
\eeq
where $^*$ denotes the complex conjugate, and $i=\sqrt{-1}$. Therefore,
${\bf R}_m$ has the eigenvalues
\beq
\lambda_j({\bf R}_m) = e^{i2\pi j/m} \quad {\rm for} \ j=0,1,\cdots,m-1.  
\label{eigenper}
\eeq
The result in (\ref{eigenqkb}) then follows. $\Box$ 

\medskip

{\bf Theorem 5} \quad 

\beq
N_{ST}(G[(K_r)_m,jn]) = 3^{(r-1)m}r^{rm-2}m \ .
\label{nstkb}
\eeq

\medskip

{\bf Proof} \quad  From (\ref{muprod}) we obtain the result 
\beqs
N_{ST}(G[(K_r)_m,jn]) & = & \frac{1}{rm}(3r)^{(r-1)m}\prod _{j=1}^{m-1}
2r[1-\cos(2\pi j/m)] \cr\cr
& = & \frac{1}{r^2m}[r(3r)^{r-1}]^m \prod_{j=1}^{m-1}4\sin^2(j\pi/m) 
\label{calc1}
\eeqs
To evaluate the product, we use the relation 
\beq
2^{m-1} \prod _{j=0}^{m-1}\sin(\epsilon+\frac{j\pi}{m}) = \sin(m\epsilon)
\label{sin}
\eeq
taking the limit $\epsilon \to 0$ and applying L'Hospital's rule.  This yields
$N_{ST}(G[(K_r)_m,jn])= (r^2m)^{-1}[r(3r)^{r-1}]^m m^2$, which, in turn, gives
the result in (\ref{nstkb}). $\Box$ 

\medskip

{\bf Corollary 5} \quad  Consider the number of spanning trees on the graph 
$G[(K_r)_m,jn]$ and take the limit $m \to \infty$.  The quantity 
measuring the growth rate in this limit is given by
\beq
\exp(z(G[(K_r)_\infty,jn])) = 3^{1-\frac{1}{r}} \ r \ . 
\label{zstkb}
\eeq

\medskip

{\bf Proof} \quad This follows from (\ref{nstkb}) and the
definition (\ref{zg}). $\Box$

\medskip

For the first nontrivial case, $r=2$, eq. (\ref{zstkb}) yields
$\exp(z(G[(K_2)_\infty,jn]))=2\sqrt{3} \simeq 3.46410$, in agreement with our
result in eq. (A.16) of \cite{ka}.

For large $r$, the ratios $R_i$, $i=1,2$ both approach unity: 
\beqs
R_1(G[(K_r)_\infty,jn]) & = & 1 + \biggl (\frac{1}{3}-\ln 3 \biggr )r^{-1} 
\cr\cr 
& & +\biggl (\frac{1}{9}-\frac{1}{3}\ln 3+\frac{1}{2}(\ln 3)^2 \biggr )r^{-2}
+ O(r^{-3}) \cr\cr
& \simeq  & 1-0.76528 \ r^{-1} + 0.34838 \ r^{-2} + O(r^{-3}) 
\label{grimratio_jn_taylor}
\eeqs
\beqs
R_2(G[(K_r)_\infty,jn]) & = & 1 + \biggl (\frac{1}{2}-\ln 3 \biggr )r^{-1}
\cr\cr
& & +\biggl (\frac{7}{24}-\frac{1}{2}\ln 3+\frac{1}{2}(\ln 3)^2 \biggr )r^{-2}
+ O(r^{-3}) \cr\cr
& \simeq  & 1-0.59861 \ r^{-1} + 0.345835 \ r^{-2} + O(r^{-3})
\label{mckayratio_jn_taylor}
\eeqs
This tendency of the growth rates to approach the upper bounds as the vertex
degree increases was also found in \cite{sw} for regular lattices.

In Table \ref{zjn} we list the growth rate (\ref{zstkb}) for $2 \le r \le 10$ 
and compare it with the upper bounds (\ref{zgrim}) and (\ref{zmckay}).  (We do
not list the lowest case, $r=1$, since $N_{ST}$ grows linearly rather than 
exponentially in $m$ for this value of $r$.) 

\begin{table}
\caption{\footnotesize{Values of $\exp(z(G[(K_r)_\infty,jn]))$, abbreviated as
$e^z$, from (\ref{zstkb}) for $2 \le r \le 10$ and comparison with the upper
bounds (\ref{zgrim}) and (\ref{zmckay}) via the ratios $R_j$, $j=1,2$ given by
(\ref{rjg}).}}
\begin{center}
\begin{tabular}{|c|c|c|c|}
\hline\hline
$r$ & $e^z$ & $R_1$ & $R_2$ \\
\hline\hline
2  & 3.464  & 0.693 & 0.786  \\ \hline 
3  & 6.240  & 0.780 & 0.838  \\ \hline
4  & 9.118  & 0.829 & 0.871  \\ \hline
5  & 12.041 & 0.860 & 0.894  \\ \hline
6  & 14.988 & 0.882 & 0.910  \\ \hline
7  & 17.950 & 0.897 & 0.921  \\ \hline
8  & 20.920 & 0.910 & 0.931  \\ \hline
9  & 23.897 & 0.919 & 0.938  \\ \hline
10 & 26.879 & 0.927 & 0.944  \\ \hline\hline
\end{tabular}
\end{center}
\label{zjn}
\end{table}

\subsection{Number of Spanning Trees in $G[(K_r)_m,id]$}

Here we calculate the number of spanning trees in the family of graphs
$G[(K_r)_m,id]$.  In \cite{lse9908} and \cite{cprsg,amcp1} studies were carried
out of the chromatic polynomials of the family $G[(K_r)_m,L]$ for general
linkage $L$, and the case of the identity linkage $L=id$ was called the
bracelet graph, $G[(K_r)_m,id] \equiv B_m(r)$.  Parenthetically, we note that
the chromatic polynomial for $G[(K_r)_m,id]$ was computed for $r=2$ in
\cite{bds}, for $r=3$ in \cite{tk}, for $r=4$ in \cite{dn} and subsequently, by
different methods, in \cite{cprsg}, and for $r=5,6$ in \cite{kb56}. The graph
$G[(K_r)_m,id]$ has $|V|=mr$, $|E|=(1/2)mr(r+1)$ and is a $\Delta$-regular
graph with uniform vertex degree $\Delta=r+1$.  For our calculation, we first
note that the definition of ${\bf a}(j,j)$ remains the same, as given in
(\ref{a0}) and
\beq 
{\bf a}(j,j+1) = {\bf a}(j+1,j) = {\bf I}_r \quad {\rm for} \ 1 \le j \le 
m \ .
\label{a2}  
\eeq

The Laplacian matrix for this the $G[(K_r)_m,id]$ graph is 
\beqs   
{\bf Q}(G[(K_r)_m,id]) &=&  
[ (r+1){\bf I}_r - {\bf a}(j,j) ] \otimes {\bf I}_m -
{\bf a}(j,j+1) \otimes {\bf R}_m - {\bf a}(j+1,j) \otimes {\bf R}^T_m
\nonumber \\
& & [ (r+2){\bf I}_r - {\bf J}_r ] \otimes {\bf I}_m - {\bf I}_r \otimes
{\bf R}_m - {\bf I}_r \otimes {\bf R}^T_m
\label{qbm}
\eeqs

\medskip

{\bf Lemma 4} \quad  The eigenvalues of ${\bf Q}(G[(K_r)_m,id])$ are 
\beq 
\lambda({\bf Q}(G[(K_r)_m,id]) = 
\cases{r + 2 - (r+e^{2i\pi j/m}+e^{-2i\pi j/m})
= 2[1-\cos(2\pi j/m)] & for $j = 0,1,\cdots,m-1$ \cr\cr  
r + 2 - (e^{2i\pi j/m}+e^{-2i\pi j/m}) = r + 2[1-\cos(2\pi j/m)] & with
multiplicity $(r-1)$.}
\label{eigenqbm}
\eeq
The proof proceeds in the same way as before, and is omitted. 

\medskip

{\bf Theorem 6} \quad  The number of spanning trees in $G[(K_r)_m,id]$ is 

\beq
N_{ST}(G[(K_r)_m,id]) = \cases{mr^{r-2} \Biggl [
(r+4)(\frac{\omega^{m/2}-\omega^{-m/2}}{\omega-\omega^{-1}})^2
\Biggr ] ^{r-1} & for even $m$
\cr\cr
mr^{r-2} (\frac{\omega^m-2+\omega^{-m}}{\omega-2+\omega^{-1}})
^{r-1} & for odd $m$.}
\label{nstbn}
\eeq
where
\beq
\omega = \frac{r+2+\sqrt{r(r+4)}}{2} \ . 
\label{omega}
\eeq

\medskip

{\bf Proof} \quad  We first use the lemma (\ref{eigenqbm}) with 
(\ref{muprod}) to obtain 
\beqs
N_{ST}(G[(K_r)_m,id]) & = & \frac{1}{rm}\Biggl [ 
\prod _{j=1}^{m-1}2[1-\cos(2\pi j/m)] \Biggr ] 
\prod_{k=0}^{m-1} \Biggl [r + 2[1-\cos(2\pi k/m)] \Biggr ] ^{r-1} \cr\cr
& = & mr^{r-2} \Biggl [ \prod _{k=1}^{m-1}[r+2-2\cos(2\pi k/m)] \Biggr
]^{r-1} \ .
\label{nstbm}
\eeqs
For even $m$, the product can be evaluated by using the identity 
\beq
\prod _{k=1}^{m-1}[x^2-2x\cos(k\pi /m)+1] = \frac{x^{2m}-1}{x^2-1}
\label{prod1}
\eeq
and setting $x=\omega$ to get 
\beqs
\prod _{k=1}^{m-1}[r+2-2\cos(2\pi k/m)] & = & (r+4) \Biggl [ 
\prod_{k=1}^{m/2-1}[r+2-2\cos(\pi k/(m/2))] \Biggr ]^2 \cr\cr
& = & (r+4) \Biggl [ \omega^{1-\frac{m}{2}} \biggl ( 
\frac{\omega^m-1}{\omega^2-1} \biggr ) \Biggr ]^2 
\cr\cr
& = & (r+4) \Biggl [ \frac{\omega^{m/2}-\omega^{-m/2}}{\omega-\omega^{-1}} 
\Biggr ] ^2
\ .
\label{prod1a}
\eeqs
For odd $m$, we use the equation
\beq 
\prod _{k=1}^{m}[\omega^2 - 2\omega\cos(2\pi k/(2m+1)) + 1] =  
\frac{\omega^{2m+1}-1}{\omega-1}
\label{prod2}
\eeq
with the same substitution (\ref{omega}) for $\omega$, to obtain
\beqs
\prod _{k=1}^{m-1}[r+2-2\cos(2\pi k/m)] & = & \Biggl [ 
\prod_{k=1}^{(m-1)/2}[r+2-2\cos(2\pi k/m)] \Biggr ]^2 \cr\cr
& = & \Biggl [ \omega^{(1-m)/2} \biggl ( \frac{\omega^m-1}{\omega-1} 
\biggr ) \Biggr ]^2 \cr\cr
& = & \frac{\omega^m-2+\omega^{-m}}{\omega-2+\omega^{-1}} \ .
\label{prod2a}
\eeqs
Combining eqs. (\ref{nstbm}), (\ref{prod1a}) and (\ref{prod2a}), we then have
the result given in (\ref{nstbn}). $\Box$ 

\medskip

We remark that the generating function of
$\frac{\omega^{m/2}-\omega^{-m/2}}{\omega-\omega^{-1}}$ for even $m$ is
$1/(1-(r+2)\zeta+\zeta^2)$, and the generating function of
$\frac{\omega^{m/2}-\omega^{-m/2}}{\omega^{1/2}-\omega^{-1/2}}$ for odd
$m$ is $(1+\zeta)/(1-(r+2)\zeta+\zeta^2)$.

\medskip

{\bf Corollary 6} \quad  Consider the number of spanning trees on the graph
$G[(K_r)_m,id]$ and take the limit $m \to \infty$.  The quantity
measuring the growth rate in this limit is given by $\omega^{1-\frac{1}{r}}$,
i.e., 
\beq
\exp(z(G[(K_r)_\infty,id])) = \Biggl [ \frac{r+2+\sqrt{r(r+4)}}{2} \ 
\Biggr ]^{1-\frac{1}{r}} \ . 
\label{zstbm}
\eeq

\medskip

{\bf Proof} \quad This follows from (\ref{nstbn}) and the
definition (\ref{zg}). $\Box$

\medskip

For $r=1$, $\exp(z(G[(K_1)_\infty,id]))=1$, reflecting the fact that 
$N_{ST}=m$ for $G[(K_1)_m,id]=C_m$, which is subexponential growth in $m$. 
For $r=2$, 
\beq
\exp(z(G[(K_2)_\infty,id])) = (2+\sqrt{3} \ )^{1/2} \simeq 1.93185
\label{zidr2}
\eeq
in agreement with our result in eq. (D.21) of \cite{a}.  For $r=3$, 
\beq
\exp(z(G[(K_3)_\infty,id])) = \biggl ( \frac{5+\sqrt{21}}{2} \ \biggr )^{2/3} 
\simeq 2.84207
\label{zidr3}
\eeq
in agreement with our result in eq. (A.88) of \cite{s3a}.

For large $r$, the ratios $R_i$, $i=1,2$ both approach unity in the manner
indicated below:
\beqs
R_1(G[(K_r)_\infty,id]) & = & 1 + (1-\ln r )r^{-1} \cr\cr
& & +\Bigl (-4-\ln r + \frac{1}{2}\ln^2 r \Bigr ) r^{-2}
+ O \biggl ( \Bigl ( \frac{\ln r}{r} \Bigr )^3 \biggr )
\label{grimratio_id_taylor}
\eeqs
\beqs
R_2(G[(K_r)_\infty,jn]) & = & 1 + \Bigl ( \frac{3}{2} - \ln r \Bigr )r^{-1}
\cr\cr
& & +\biggl( -\frac{27}{8} -\frac{3}{2}\ln r + \frac{1}{2}\ln^2 r \Bigr )r^{-2}
+ O \biggl ( \Bigl ( \frac{\ln r}{r} \Bigr )^3 \biggr )
\label{mckayratio_id_taylor}
\eeqs

In Table \ref{zid} we list the growth rate (\ref{zstbm}) for $2 \le r \le 10$
and compare it with the upper bounds (\ref{zgrim}) and (\ref{zmckay}).  As
expected, the approach of $R_1$ and $R_2$ to unity as $r$ increases is less
rapid for $G[(K_r)_\infty,id]$ than for $G[(K_r)_\infty,jn]$ because of the
fact that the family with $L=id$ has a lower vertex degree, $\Delta=r+1$ than
that of the family with $L=jn$, for which $\Delta=3r-1$.

\begin{table}
\caption{\footnotesize{Values of $\exp(z(G[(K_r)_\infty,id]))$, abbreviated as
$e^z$, from (\ref{zstbm}) for $2 \le r \le 10$ and comparison with the upper
bounds (\ref{zgrim}) and (\ref{zmckay}) via the ratios $R_j$, $j=1,2$ given by
(\ref{rjg}).}}
\begin{center}
\begin{tabular}{|c|c|c|c|}
\hline\hline
$r$ & $e^z$ & $R_1$ & $R_2$ \\
\hline\hline
2  & 1.932  & 0.644 & 0.837  \\ \hline
3  & 2.842  & 0.711 & 0.842  \\ \hline
4  & 3.751  & 0.750 & 0.851  \\ \hline
5  & 4.664  & 0.777 & 0.860  \\ \hline
6  & 5.582  & 0.797 & 0.867  \\ \hline
7  & 6.505  & 0.813 & 0.874  \\ \hline
8  & 7.433  & 0.826 & 0.879  \\ \hline
9  & 8.365  & 0.836 & 0.884  \\ \hline
10 & 9.301  & 0.846 & 0.889  \\ \hline\hline
\end{tabular}
\end{center}
\label{zid}
\end{table}

Acknowledgments.  R.S. thanks Profs. Dominic Welsh and Marc Noy for their
organization of the stimulating CRM Workshop on Tutte Polynomials. This
research was partially supported by the NSF grant PHY-0098527.

\section{Appendix}

In this appendix we list the cubic and quartic equations that yield the
$\lambda_{T,3,d,j}$'s for $T(G[(K_3)_m,jn],x,y)$ that were not already given in
the text.  The $\lambda_{T,3,1,j}$, $j=1,2,3,4$ are solutions of the quartic
equation
\beq
\xi^4+b_{43}\xi^3+b_{42}\xi^2+b_{41}\xi+b_{40}=0
\label{quarticeq}
\eeq
where
\beqs
& & b_{43} = -(y^9+3y^8+6y^7+10y^6+15y^5+3xy^3+21y^4+6xy^2+31y^3 \cr\cr
& & +3x^2+9xy+39y^2+21x+45y+36)
\label{b43}
\eeqs
\beqs
& & b_{42} = 3y(xy^{11}+4xy^{10}+y^{11}+x^2y^8+10xy^9+5y^{10}+3x^2y^7+23xy^8 
\cr\cr
& & +15y^9+6x^2y^6+39xy^7+31y^8+10x^2y^5+53xy^6+42y^7+14x^2y^4 \cr\cr
& & +62xy^5+41y^6+3x^3y^2+17x^2y^3+65xy^4+28y^5+3x^3y+30x^2y^2 \cr\cr
& & +66xy^3+8y^4+21x^2y+67xy^2-13y^3-6x^2+6xy-38y^2-30x-66y-36) \cr\cr
& & 
\label{b42}
\eeqs
\beqs
& & b_{41}=-9y^3(y+1)(xy^{11}+x^3y^8+5xy^{10}+2x^3y^7+4x^2y^8+16xy^9+2x^3y^6
 \cr\cr
& & +7x^2y^7+38xy^8+x^3y^5+7x^2y^6+57xy^7-2y^8+9x^2y^5+72xy^6-11y^7 \cr\cr
& & +14x^2y^4+81xy^5-23y^6+3x^3y^2+20x^2y^3+73xy^4-41y^5+6x^3y+26x^2y^2 \cr\cr
& & +44xy^3-65y^4-3x^3+3x^2y-84y^3-18x^2-48xy-85y^2-33x-61y-18) \cr\cr
& & 
\label{b41}
\eeqs
\beqs
& & b_{40}=27y^6(y+1)^2(y^2+y+1)(xy-1)(x^2y^4+2x^2y^3+2xy^4-x^2y^2 \cr\cr
& & -3xy^3-y^4-5xy^2-3y^3+2x+4y+2) \ .
\label{b40}
\eeqs

The $\lambda_{T,3,0,j}$, $j=2,3,4$, solutions of the cubic equation
\beq
\xi^3+b_{32}\xi^2+b_{31}\xi+b_{30}=0 
\label{cubiceq}
\eeq
where
\beqs
& & b_{32}=-(y^9+3y^8+6y^7+10y^6+15y^5+3xy^3+21y^4+x^3+6xy^2+28y^3 \cr\cr
& & +9x^2+16xy+36y^2+26x+38y+24)
\label{b32}
\eeqs
\beqs
& & b_{31}=y(3xy^{11}+x^3y^8+12xy^{10}+3x^3y^7+9x^2y^8+36xy^9+3y^{10} \cr\cr
& & +6x^3y^6+27x^2y^7+90xy^8+12y^9+10x^3y^5+54x^2y^6+153xy^7+17y^8 \cr\cr
& & +15x^3y^4+84x^2y^5+194xy^6-6y^7+3x^4y^2+21x^3y^3+102x^2y^4+184xy^5 \cr\cr
& & -62y^6+6x^4y+40x^3y^2+96x^2y^3+117xy^4-134y^5-3x^4+18x^3y+75x^2y^2 \cr\cr
& & +16xy^3-192y^4-24x^3-36x^2y-85xy^2-211y^3-69x^2-156xy-189y^2 \cr\cr
& & -84x-120y-36) 
\label{b31}
\eeqs
\beqs
& & b_{30}=-3y^4(y+1)(x^4y^7+3x^4y^6+6x^3y^7+3x^4y^5+7x^3y^6+3x^2y^7 \cr\cr
& & +x^4y^4-6x^3y^5-15x^2y^6-4xy^7-3x^4y^3-15x^3y^4-31x^2y^5-12xy^6 \cr\cr
& & -11x^3y^3-14x^2y^4+7xy^5+5y^6+x^4y+9x^3y^2+21x^2y^3+33xy^4+9y^5 \cr\cr
& & +6x^3y+25x^2y^2+28xy^3+y^4-2x^3-x^2y-4xy^2-11y^3-6x^2-14xy \cr\cr
& & -12y^2-4x-4y) \ .
\label{b30}
\eeqs

\vfill
\eject

\end{document}